\newif\ifdraft
\draftfalse
\newif\ifcameraready
\camerareadytrue

\documentclass[camera,letterpaper,nomarginnotes,nonarrowgutter]{jpaper}

\usepackage[bookmarks=true,breaklinks=true,colorlinks,linkcolor=black,citecolor=NavyBlue,urlcolor=blue]{hyperref}

\usepackage{algorithmic}
\usepackage{graphicx}
\usepackage{textcomp}
\usepackage[dvipsnames]{xcolor}

\usepackage{titlecaps}

\usepackage{balance}
\usepackage{textcomp}
\usepackage{pifont}

\usepackage{booktabs}
\usepackage{glossaries-extra} 
\setkeys{glslink}{hyper=false} 
\usepackage{setspace}
\usepackage{listings}
\usepackage[linesnumbered]{algorithm2e}
\usepackage{siunitx} 
\usepackage{subcaption}
\usepackage{tikz}
\usepackage{multirow}
\usepackage{enumitem}
\usepackage[utf8]{inputenc}

\usepackage{titlesec}

\usepackage{interval} 
\usepackage{duckuments} 
\usepackage{algorithmic} 

\usepackage[us,12hr]{datetime}
\usepackage[en-GB, useregional=numeric]{datetime2}
\usepackage{fourier-orns}
\usepackage{fancyhdr}
\usepackage{pgfornament}
\usetikzlibrary{calc}
\usepackage{url}
\usepackage[sort,compress]{cite}

\def\UrlBreaks{\do\/\do-\/\do.\/\do:}

\expandafter\def\expandafter\UrlBreaks\expandafter{\UrlBreaks
  \do\a\do\b\do\c\do\d\do\e\do\f\do\g\do\h\do\i\do\j
  \do\k\do\l\do\m\do\n\do\o\do\p\do\q\do\r\do\s\do\t
  \do\u\do\v\do\w\do\x\do\y\do\z\do\A\do\B\do\C\do\D
  \do\E\do\F\do\G\do\H\do\I\do\J\do\K\do\L\do\M\do\N
  \do\O\do\P\do\Q\do\R\do\S\do\T\do\U\do\V\do\W\do\X
  \do\Y\do\Z}

\newcommand{\revdel}[1]{}
\newcommand{\X}{PuDHammer}

\newcommand*\DRAMTIMING[1]{t\textsubscript{#1}}
\newcommand*\nCHIPS{316}
\newcommand*\nMODULES{40}

\newcommand\src{$src$}
\newcommand\dst{$dst$}
 \newcommand*\head[1]{\noindent\textbf{#1}}

\definecolor{gfored}{rgb}{0.580, 0.050, 0.211}
\definecolor{ao}{rgb}{0.007, 0.520, 0.867}
\definecolor{moegi}{rgb}{0.357, 0.537, 0.188}
\definecolor{jl}{rgb}{1.0, 0.2, 0.8}
\definecolor{brown(web)}{rgb}{0.65, 0.16, 0.16}
\definecolor{bisque}{rgb}{1.0, 0.89, 0.77}
\definecolor{nbs}{rgb}{0.88, 0.07, 0.37}
\definecolor{yt}{rgb}{0.58, 0.44, 0.86}
\definecolor{iy}{rgb}{0.0, 0.36, 0.05}
\definecolor{mel}{rgb}{0.9, 0.55, 0.31}
\definecolor{ouscolor}{rgb}{0.0, 0.2, 0.4}

\newcommand{\dingOne}{\circledtest{1}}
\newcommand{\dingTwo}{\circledtest{2}}
\newcommand{\dingThree}{\circledtest{3}}

\newcommand{\one}{1)}
\newcommand{\two}{2)}
\newcommand{\three}{3)}

\newcommand{\five}{5)}

\newcommand{\xxx}[1]{\param{XXX}} 

\newcommand{\ignore}[1]{}

\newcommand{\om}[1]{{#1}}
\newcommand{\agyi}[1]{{#1}}

\newcommand{\cref}[1]{\hyperref[ref:#1]{#1}}

\newcommand{\param}[1]{\textcolor{red}{#1}}

\ifdraft
    \usepackage[colorinlistoftodos,prependcaption,textsize=scriptsize]{todonotes} 
    \paperwidth=\dimexpr \paperwidth + 4cm\relax
    \oddsidemargin=\dimexpr\oddsidemargin + 2cm\relax
    \evensidemargin=\dimexpr\evensidemargin + 2cm\relax
    \marginparwidth=\dimexpr \marginparwidth + 2cm\relax

    \newcommand{\agycomment}[1]{\todo[size=\scriptsize, linecolor=orange, bordercolor=orange, backgroundcolor=white]{\textcolor{gfored}{\textbf{@gy:} #1}}}

    \newcommand{\atb}[1]{\textcolor{ao}{#1}}

    \newcommand{\atbcomment}[1]{\todo[size=\scriptsize, linecolor=orange, bordercolor=orange, backgroundcolor=white]{\textcolor{ao}{\textbf{@atb:} #1}}}

    \newcommand{\nbcomment}[1]{\todo[size=\scriptsize, linecolor=orange, bordercolor=orange, backgroundcolor=white]{\textcolor{nbs}{\textbf{@nb:} #1}}}

    \newcommand{\ous}[1]{\textcolor{ouscolor}{#1}}
    \newcommand{\ouscomment}[1]{\todo[size=\scriptsize, linecolor=orange, bordercolor=orange, backgroundcolor=white]{\textcolor{ouscolor}{\textbf{@ous:} #1}}}

    \newcommand{\hluo}[1]{\textcolor{moegi}{#1}}
    \newcommand{\hluocomment}[1]{\todo[size=\scriptsize, linecolor=orange, bordercolor=orange, backgroundcolor=white]{\textcolor{moegi}{\textbf{@hluo:} #1}}}

    \newcommand{\iey}[1]{\textcolor{iy}{#1}}
    \newcommand{\ieycomment}[1]{\todo[size=\scriptsize, linecolor=orange, bordercolor=orange, backgroundcolor=white]{\textcolor{iy}{\textbf{@iey:} #1}}}

    \newcommand{\om}[1]{\textcolor{teal}{#1}}

\else
    \renewcommand{\param}[1]{\textcolor{black}{#1}}

    \newcommand{\agy}[1]{{#1}}
    \newcommand{\agycomment}[1]{}
    \newcommand{\agyinline}[1]{}

    \newcommand{\mscomment}[1]{}
    
    \newcommand{\atb}[1]{{#1}}
    \newcommand{\atbcomment}[1]{}

    \newcommand{\nbcomment}[1]{}

    \newcommand{\hluo}[1]{{#1}}
    \newcommand{\hluocomment}[1]{}

    \newcommand{\iey}[1]{#1}
    \newcommand{\ieycomment}[1]{}

    \newcommand{\ous}[1]{#1}
    \newcommand{\ouscomment}[1]{}
    
\fi

\definecolor{frenchblue}{rgb}{0.19, 0.55, 0.91}

\usepackage[most]{tcolorbox} 
\tcbset{before skip=4.1pt, after skip=4.1pt}

\newtcolorbox[auto counter]{obsx}[3][]{%
    colframe = #2!45,
    colback  = #2!30,
    colbacktitle=#2!20,
    coltitle=black,
    fonttitle=\bfseries, 
    title=#3~\thetcbcounter.\ ,
    enhanced,
    attach boxed title to top left={yshift=-2.8mm, xshift=0.15cm},
    bottom=-2.2pt,
    #1%
}

\usepackage[most]{tcolorbox} 
\newtcolorbox[auto counter]{tkx}[2][]{%
    enhanced, breakable, center title,
    colframe = #2!55,
    colback  = #2!15,
    colbacktitle=#2!20,
    left=-0.5pt,
    right=-0.5pt,
    bottom=-2pt,
    top=-0.25pt,
    #1%
}
\newtcolorbox[auto counter]{obssx}[2][]{%
    enhanced, breakable, center title,
    colframe = #2!65,
    colback  = #2!15,
    colbacktitle=#2!20,
    left=-0.5pt,
    right=-0.5pt,
    bottom=-2pt,
    top=-0.25pt,
    #1%
}

\newcounter{obs}
\newcommand\observation[1]{
\refstepcounter{obs}
\begin{obssx}{DarkOrchid}
\noindent\textbf{Observation~\theobs.} #1
\end{obssx}
}

\newcounter{tkw}
\newcommand\takeaway[1]{
\stepcounter{tkw}
\begin{tkx}{Sepia}
\noindent\textbf{Takeaway~\thetkw.} #1
\end{tkx}
}

\newcommand{\figref}[1]{Fig.~\ref{#1}}

\newcommand{\secref}[1]{§\ref{#1}}

\newcommand*\circledtest[1]{\tikz[baseline=(char.base)]{
            \node[shape=circle,fill,inner sep=0.3pt] (char) {\textcolor{white}{#1}};}}

\makeatletter
\g@addto@macro{\normalsize}{%
 \setlength{\abovedisplayskip}{2pt plus 1pt minus 1pt}
 \setlength{\belowdisplayskip}{1pt plus 1pt minus 1pt}
  \setlength{\abovedisplayshortskip}{1pt}
  \setlength{\belowdisplayshortskip}{1pt}
  \setlength{\intextsep}{1pt plus 1pt minus 1pt}
  \setlength{\textfloatsep}{2pt plus 1pt minus 1pt}
  \setlength{\skip\footins}{3pt plus 1pt minus 1pt}}
 \titlespacing\section{2pt}{2pt plus 2pt minus 2pt}{2pt plus 2pt minus 2pt}
 \titlespacing\subsection{2pt}{2pt plus 2pt minus 2pt}{2pt plus 2pt minus 2pt}
\makeatother

\ifcameraready

    \newcommand{\atbcr}[2]{\ifnum#1>-1\textcolor{black}{#2}\else{#2}\fi}
    \newcommand{\ieycr}[2]{\ifnum#1>-1\textcolor{black}{#2}\else{#2}\fi}
    \newcommand{\omcr}[2]{\ifnum#1>-1\textcolor{black}{#2}\else{#2}\fi}
    \newcommand{\omcrcomment}[1]{}
    \newcommand{\crdiscussion}[2]{}{}

    \newcommand{\ominline}[1]{}
    \newcommand{\ieycrcomment}[1]{}
    \newcommand{\ieyinline}[1]{}
\else
\usepackage[colorinlistoftodos,prependcaption,textsize=scriptsize]{todonotes} 
    \paperwidth=\dimexpr \paperwidth + 4cm\relax
    \oddsidemargin=\dimexpr\oddsidemargin + 2cm\relax
    \evensidemargin=\dimexpr\evensidemargin + 2cm\relax
    \marginparwidth=\dimexpr \marginparwidth + 2cm\relax

    \newcommand{\atbcr}[2]{\ifnum#1=\value{version}\textcolor{ao}{#2}\else{#2}\fi}

    \newcommand{\ieycr}[2]{\ifnum#1=\value{version}\textcolor{blue}{#2}\else{#2}\fi}

    \newcommand{\ieycrcomment}[1]{\todo[linecolor=orange, bordercolor=orange, backgroundcolor=white]{\textcolor{iy}{\textbf{@Ismail:} #1}}}
    \newcommand{\ieyinline}[1]{\\\textcolor{iy}{\textbf{@Ismail:} #1}}

    \newcommand{\crdiscussion}[2]{\omcrcomment{#1\\\textcolor{blue}{\textbf{@Ismail:}#2}}}
    \newcommand{\omcr}[2]{\ifnum#1=\value{version}\textcolor{red}{#2}\else{#2}\fi}

    \newcommand{\omcrcomment}[1]{\todo[linecolor=orange, bordercolor=orange, backgroundcolor=white]{\textcolor{red}{\textbf{@Onur:} #1}}}
    \newcommand{\ominline}[1]{\\\textcolor{red}{\textbf{@Onur:} #1}}

\fi

\setabbreviationstyle[acronym]{long-short}

\newcommand{\tras}[0]{t_{RAS}}
\newcommand{\trp}[0]{t_{RP}}
\newcommand{\trc}[0]{t_{RC}}
\newcommand{\trefi}[0]{t_{REFI}}
\newcommand{\trefw}[0]{t_{REFW}}

\newcommand{\vdd}[0]{\texttt{$V_{DD}$}}

\newcommand{\pum}[0]{PuM}
\newcommand{\pim}[0]{PiM}
\newcommand{\pnm}[0]{PnM}

\newcommand{\pud}[0]{{PuD}}

\newcommand{\act}[0]{\texttt{ACT}}
\newcommand{\pre}[0]{\texttt{PRE}}
\newcommand{\refresh}[0]{REF}
\newcommand{\wri}[0]{\texttt{WR}}
\newcommand{\rd}[0]{\texttt{RD}}

\newcommand{\cots}[0]{COTS}
\newcommand{\comra}[0]{CoMRA}
\newcommand{\simra}[0]{SiMRA}

\newacronym{iqr}{$IQR$}{inter-quartile range}
\newacronym{act}{\act{}}{activate}
\newacronym{pre}{\pre{}}{precharge}
\newacronym{ref}{\refresh{}}{refresh}
\newacronym{wr}{\wri{}}{write}
\newacronym{rd}{\rd{}}{read}
\newacronym{pim}{\pim{}}{Processing-in-Memory}
\newacronym{pnm}{\pnm{}}{Processing-near-Memory}

\newacronym{pum}{\pum{}}{Processing-using-Memory}
\newacronym{pud}{\pud{}}{Processing-using-DRAM}

\newacronym{cots}{\cots{}}{commercial off-the-shelf}
\newacronym{comra}{\comra{}}{}
\newacronym{comraL}{which we call \comra{}}{\underline{co}nsecutive \underline{m}ultiple-\underline{r}ow \underline{a}ctivation}
\newacronym{simraL}{which we call \simra{}}{\underline{si}multaneous \underline{m}ultiple-\underline{r}ow \underline{a}ctivation}
\newacronym{simra}{\simra{}}{}

\newacronym{jedec}{JEDEC}{Joint Electron Device Engineering Council}

\newcommand{\hcfirst}[0]{$HC_{first}$}
\newacronym{hcfirst}{\hcfirst{}}{the minimum {hammer count} {required to induce the first bitflip}} 
\newcommand{\ber}[0]{$BER$}
\newacronym{ber}{\ber{}}{bit error rate}
\newacronym{wcdp}{$WCDP$}{worst-case data pattern}
\newacronym{taggon}{$t_{AggOn}$}{the time that an aggressor row stays active}

\newacronym{trefw}{$\trefw$}{}
\newacronym{tras}{$\tras$}{}
\newacronym{trp}{$\trp$}{}
\newacronym{trefi}{$\trefi$}{}

\newcommand{\rhmemisolationrefs}[0]{\cite{fournaris2017exploiting, poddebniak2018attacking, tatar2018throwhammer, carre2018openssl, barenghi2018software, zhang2018triggering, bhattacharya2018advanced, google-project-zero, kim2014flipping, rowhammergithub, seaborn2015exploiting, van2016drammer, gruss2016rowhammer, razavi2016flip, pessl2016drama, xiao2016one, bosman2016dedup, bhattacharya2016curious, burleson2016invited, qiao2016new, brasser2017can, jang2017sgx, aga2017good, mutlu2017rowhammer, tatar2018defeating, gruss2018another, lipp2018nethammer, van2018guardion, frigo2018grand, cojocar2019eccploit,  ji2019pinpoint, mutlu2019rowhammer, hong2019terminal, kwong2020rambleed, frigo2020trrespass, cojocar2020rowhammer, weissman2020jackhammer, zhang2020pthammer, yao2020deephammer, deridder2021smash, hassan2021utrr, jattke2022blacksmith, tol2022toward, kogler2022half, orosa2022spyhammer, zhang2022implicit, liu2022generating, cohen2022hammerscope, zheng2022trojvit, fahr2022frodo, tobah2022spechammer, rakin2022deepsteal, park2016statistical, park2016experiments,lim2017active, ryu2017overcoming, yun2018study, yang2019trap, walker2021ondramrowhammer, kim2020revisiting, orosa2021deeper, yaglikci2022understanding, khan2018analysis, agarwal2018rowhammer, li2014write, ni2018write, genssler2022reliability, mutlu2023fundamentally}}

\newcommand{\pudAllCitations}[0]{\cite{seshadri2013rowclone,seshadri2018rowclone,chang2016low,olgun2022pidram, seshadri2015fast,seshadri2019dram,seshadri2016processing,seshadri2017simple,seshadri2016buddy,gao2022frac,xin2020elp2im,besta2021sisa, li2017drisa, deoliveira2024mimdram, yuksel2024functionally, hajinazar2021simdram, gao2019computedram, seshadri2017ambit, deng2018dracc, angizi2019graphide, li2018scope, ferreira2022pluto,deng2019lacc,sutradhar2021look,sutradhar2020ppim, Li2018SCOPEAS, olgun2021quac, kim2019drange, kim2018dram, olgun2023dram, oliveira2022accelerating, deoliveira2024mimdram, oliveira2025proteus, mutlu2024memory, mutlu2025memory}}

\newcommand{\dataMovementProblemsCitations}[0]{\cite{mutlu2013memory,mutlu2015research,dean2013tail,kanev_isca2015,ferdman2012clearing,wang2014bigdatabench,mutlu2019enabling,mutlu2019processing,mutlu2020intelligent,ghose.ibmjrd19,mutlu2020modern,oliveira2021damov,boroumand2018google,boroumand2021google,wang2016reducing, pandiyan2014quantifying,koppula2019eden,kang2014co,mckee2004reflections,wilkes2001memory,kim2012case,wulf1995hitting,ghose.sigmetrics20,ahn2015scalable,PEI,hsieh2016transparent,wang2020figaro,sites1996}}

\begin{document}

\title{PuDHammer: Experimental Analysis of Read Disturbance Effects of Processing-using-DRAM in Real DRAM Chips}
\author{
{{\.I}smail Emir Y{\"u}ksel}\qquad
{Akash Sood}\qquad
{Ataberk Olgun}\qquad
{O\u{g}uzhan Canpolat}\qquad
{Haocong Luo}\\
{F. Nisa Bostanc{\i}}\qquad
{Mohammad Sadrosadati}\qquad
{A.~Giray~Ya\u{g}l{\i}k\c{c}{\i}}\qquad
{Onur Mutlu}
\vspace{0mm}\\\\
{ETH Z{\"u}rich}
}


\maketitle

    \renewcommand{\headrulewidth}{0pt}
    \fancypagestyle{firstpage}{
        \fancyhead{} 
        \fancyhead[C]{
      } 
    \renewcommand{\footrulewidth}{0pt}
    }
  \thispagestyle{firstpage}

\pagenumbering{arabic}

\newcounter{version}
\setcounter{version}{999}
\glsresetall
\begin{abstract}
\gls{pud} is a promising \omcr{0}{paradigm} for alleviating the data movement bottleneck using a DRAM array's massive internal parallelism and bandwidth to execute very wide data-parallel operations. Performing a \gls{pud} operation involves activating \ieycr{1}{multiple} DRAM rows 
in quick succession or simultaneously, i.e., multiple-row activation.
Multiple-row activation is fundamentally different from conventional memory access patterns that activate one DRAM row at a time.
However, repeatedly activating \emph{even one} DRAM row (e.g., RowHammer) can induce bitflips in unaccessed DRAM rows because modern DRAM is subject to read disturbance, a worsening safety, security, and reliability issue.
Unfortunately, no prior work investigates the effects of multiple-row activation, as commonly used by \gls{pud} operations, on DRAM read disturbance.

\ieycr{0}{In this paper,}
we present the first characterization study of read disturbance effects of multiple-row activation-based \gls{pud} (which we call \X{}) using \nCHIPS{} real DDR4 DRAM chips from four major \ieycr{0}{DRAM} manufacturers. 
Our detailed characterization results covering various operational conditions and parameters (i.e., temperature, data patterns, access patterns, timing parameters, and spatial variation) show that
\one{}~\ieycr{0}{\X{}} significantly exacerbates the read disturbance vulnerability,
\atb{causing up to 158.58$\times{}$ reduction in \gls{hcfirst}, \ieycr{0}{compared to RowHammer}},
\two{}~\ieycr{0}{\X{}} is affected by various operational conditions and parameters, 
\three{}~combining RowHammer with \ieycr{0}{\X{}} is more effective
than using RowHammer alone to induce read disturbance errors \ieycr{0}{(e.g., compared to RowHammer,} doing so
reduces \gls{hcfirst} by 1.66$\times{}$ on average across all tested rows\ieycr{0}{), and 
4) \ieycr{0}{\X{}} bypasses an in-DRAM RowHammer mitigation mechanism called Target Row Refresh and induces more bitflips than RowHammer.}

To develop future robust \gls{pud}-enabled systems in the presence of \ieycr{0}{\X{}}, we 1) \omcr{1}{develop} three countermeasures and 2) adapt and evaluate the effectiveness of state-of-the-art RowHammer mitigation standardized by industry, called Per Row Activation Count\omcr{0}{ing} (PRAC). We show that \ieycr{0}{the adapted PRAC} incur\ieycr{0}{s large} performance overheads to mitigate \ieycr{0}{\X{}} \ieycr{0}{(e.g., an average performance overhead of 48.26\% across 60 five-core multiprogrammed workloads).} We hope and expect that our findings motivate and guide system-level and architectural solutions to enable read-disturbance-resilient future PuD systems.
\end{abstract}

\glsresetall

\section{Introduction}

Modern computing systems move vast amounts of data between main memory (DRAM) and processing elements (e.g., CPU and GPU)~\cite{mutlu2019processing, mutlu2020modern}. Unfortunately, this data movement is a major bottleneck that consumes a large fraction of execution time and energy in many modern applications~\dataMovementProblemsCitations{}. 
\gls{pud}~\ieycr{0}{\cite{seshadri2017ambit,seshadri2015fast,seshadri2019dram,hajinazar2021simdram,mutlu2024memory,mutlu2025memory}} is a promising paradigm that can alleviate the data movement bottleneck. \gls{pud} uses the analog operational properties of the DRAM \hluo{array circuitry} to enable massively parallel in-DRAM computation (i.e., \gls{pud} operations), which can be used \hluo{to accelerate} important applications including databases and web search~\ieycr{0}{\cite{chan1998bitmap,o2007bitmap,li2014widetable,li2013bitweaving,goodwin2017bitfunnel,seshadri2013rowclone,seshadri2017ambit,seshadri2015fast,hajinazar2021simdram,wu2005fastbit,wu1998encoded,redis-bitmaps}, data analytics~\cite{perach2023understanding,seshadri2017ambit,jun2015bluedbm,torabzadehkashi2019catalina,lee2020smartssd,besta2021sisa}, graph processing~\cite{beamer2012direction,besta2021sisa,li2016pinatubo,gao2021parabit,hajinazar2021simdram}, genome analysis~\cite{alser2017gatekeeper,loving2014bitpal,xin2015shifted,cali2020genasm,kim2017grim,myers1999fast}, cryptography~\cite{han1999optical,tuyls2005xor}, hyper-dimensional computing~\cite{kanerva1992sparse,kanerva2009hyperdimensional,karunaratne2020memory}, and generative AI~\cite{he2025papi,gu2025pim, zhou2022transpim, park2024attacc,seo2024ianus,yun2024duplex,heo2024neupims,brown2020language,devlin2019bert,goodfellow2014generative}.}

A wide variety of \gls{pud} operations (e.g., in-DRAM data copy and bulk bitwise operations) rely on a key \gls{pud} technique called \emph{multiple-row activation}\omcr{0}{,} which accesses (activates) multiple DRAM rows in quick succession or simultaneously~\cite{seshadri2017ambit,seshadri2015fast,seshadri2019dram,hajinazar2021simdram,gao2019computedram,gao2022frac,olgun2023dram,olgun2021quac,olgun2022pidram, yuksel2024functionally, yuksel2024simultaneous,deoliveira2024mimdram, seshadri2016buddy, seshadri2016processing, seshadri2013rowclone, seshadri2018rowclone,jahshan2024majork}. 
Multiple-row activation is fundamentally different \omcr{0}{from} conventional DRAM operations that access only a \emph{single} DRAM row at a time.

Unfortunately, with aggressive technology node scaling, \omcr{0}{repeatedly} accessing \omcr{0}{even} a single DRAM row disturbs the data integrity of \emph{unaccessed} physically-adjacent DRAM rows and causes bitflips~\rhmemisolationrefs{}. \ieycr{0}{\emph{RowHammer}~\cite{kim2014flipping} and \emph{RowPress}~\cite{luo2023rowpress} are two prominent examples of DRAM read disturbance phenomena where a DRAM row (i.e., victim row) can experience bitflips when a nearby DRAM row (i.e., aggressor row) is 1)~repeatedly activated (i.e., hammered)~\cite{kim2014flipping,mutlu2019rowhammer,mutlu2023fundamentally,mutlu2025memory} or 2)~kept open for a long period (i.e., pressed)~\cite{orosa2021deeper, luo2023rowpress,luo2024experimental,mutlu2025memory}}.~\omcr{0}{Unfortunately}, no prior work explores the read disturbance effects of multiple-row activation-based \gls{pud} operations.

In this paper, we present the first experimental characterization of the read disturbance effects of \omcr{0}{\emph{multiple-row activation-based \gls{pud} operations}}, \ieycr{0}{\X{}}, on \nCHIPS{} \gls{cots} DDR4 DRAM chips from four major DRAM manufacturers \omcr{0}{(in \nMODULES{} DRAM modules)}. We characterize the read disturbance effect of two types of multiple-row activation: \one{} \gls{comraL}, used \omcr{0}{for} in-DRAM data copy~\cite{seshadri2017simple,seshadri2013rowclone,seshadri2018rowclone,gao2019computedram,gao2022frac,yuksel2024simultaneous,olgun2022pidram} (\secref{sec:comra}) and \two{} \gls{simraL}, used \omcr{0}{for} in-DRAM bitwise operations~\cite{seshadri2015fast,seshadri2017ambit,seshadri2016buddy,seshadri2016processing,seshadri2017simple,gao2019computedram,gao2022frac,olgun2021quac,yuksel2024functionally,yuksel2024simultaneous} (\secref{sec:simra}) \omcr{0}{under various} operational conditions and parameters (i.e., data patterns, temperature, access patterns, timing parameters, and spatial variation). We also analyze 1) how combining RowHammer with \ieycr{0}{\X{}} (i.e., a combined pattern that performs RowHammer and \ieycr{0}{\X{}} repeatedly) affects read disturbance (\secref{sec:combined}) and 2) the effectiveness of \omcr{0}{an} in-DRAM mitigation mechanism, broadly referred to as Target Row Refresh (TRR)~\cite{hassan2021utrr,frigo2020trrespass,micron2016trr,zhang2022softtrr,marazzi2022protrr}, against \ieycr{0}{\X{}} (\secref{sec:trr}).

Based on our characterization, we make \param{26} new empirical observations and share \param{9} key takeaway lessons. We highlight four of our major new results. 
First, repeatedly performing multiple-row activation \omcr{0}{greatly} increases the DRAM chip's read disturbance vulnerability. We find that both \comra{} and \simra{} decrease \gls{hcfirst} in all tested DRAM chips from four manufacturers. We observe that the lowest \gls{hcfirst} observed \omcr{1}{due to} \comra{} and \simra{} are \ieycr{0}{13.98$
\times$} and \ieycr{0}{158.58$
\times$} lower than the lowest \gls{hcfirst} observed \omcr{1}{due to} RowHammer, respectively. 
Second, operational conditions and parameters impact \X{} (especially \simra{}). We find that hammering with \simra{} is significantly affected by data pattern \omcr{1}{and} row on time \omcr{1}{(i.e}., RowPress~\cite{luo2023rowpress}) and can change the average \gls{hcfirst} of victim rows in a DRAM chip by up to \ieycr{0}{57.80$\times$} \omcr{1}{and} \ieycr{0}{270.27$\times$} \ieycr{0}{across all tested data patterns \omcr{1}{and} row on time values.} Third, a combined RowHammer and \X{} \ieycr{1}{access} pattern is much more effective \omcr{0}{at} inducing read disturbance bitflips than using RowHammer alone. We observe that, compared to RowHammer, the average \gls{hcfirst} of victim rows in a DRAM chip decreases by up to \ieycr{0}{1.34$\times$}, \ieycr{0}{1.22$\times$} when RowHammer is combined with \comra{} or \simra{}, respectively. We find that a combination of RowHammer, \comra{}, and \simra{} is the most effective \ieycr{1}{access} pattern among \ieycr{0}{the} tested patterns and reduces average \gls{hcfirst} of victim rows in a DRAM chip by \ieycr{0}{1.66$\times$}. Fourth, we observe that in a tested SK Hynix DDR4 DRAM module, both \comra{} and \simra{} bypass the TRR mechanism and induce more bitflips than RowHammer. For example, \simra{} and \comra{} respectively induce 11340$\times$ and 1.10$\times$ more bitflips than RowHammer on average in the presence of TRR.

\ieycr{0}{Our characterization results suggest that future \gls{pud}-enabled systems
should take \X{} into account to maintain the fundamental robustness property of memory isolation. Based on our findings, to mitigate \X{}, we} 1) \omcr{1}{develop} and qualitatively analyze \param{three} countermeasures \ieycr{0}{against \X{} \omcr{1}{that} modify the DRAM chips and DRAM interface (\secref{sec:countermeasures})} and 2) adapt and evaluate the state-of-the-art RowHammer solution standardized by industry, called Per Row Activation Count\ieycr{1}{ing} (PRAC)~\cite{saroiu2024ddr5, canpolat2025chronus,canpolat2024understanding,kim2023ddr5,jedec2024jesd795c} (\secref{subsec:rh_mitigation}). We find that \omcr{0}{the adapted} PRAC \omcr{0}{solution to \X{}} incurs an average system performance overhead of 48.26\% across all tested 60 five-core multiprogrammed workload mixes.

This paper makes the following key contributions:
\begin{itemize}
    \item To our knowledge, this is the first work to analyze and experimentally demonstrate the interaction between read disturbance and \omcr{0}{Processing-using-DRAM (PuD)} operations in \gls{cots} \omcr{0}{DRAM} chips. We extensively characterize the read disturbance effect of multiple-row activation-based \gls{pud} operations on \nCHIPS{} chips from \nMODULES{} real DRAM modules. 
    \item Our results show that \one{} multiple-row activation \omcr{0}{greatly} amplif\omcr{0}{ies} the DRAM read disturbance errors \ieycr{0}{across all tested manufacturers (e.g., up to 158.58$\times$ reduction in \gls{hcfirst})}, \two{} the read disturbance effect of multiple-row activation \omcr{1}{depends on} operational conditions and parameters with \omcr{1}{large} variations in some cases \ieycr{0}{(e.g., up to 270.27$\times$ change in \gls{hcfirst})}, \three{} combining RowHammer with multiple-row activation is more effective than using RowHammer alone for inducing the first read disturbance error \ieycr{0}{(e.g., 1.66$\times$ reduction in \gls{hcfirst} on average), and 4) \X{} bypasses the in-DRAM TRR mechanism\omcr{1}{,} induc\omcr{1}{ing many} more bitflips in a DRAM row than RowHammer does, in the presence of TRR (e.g., 11340$\times$ more bitflips on average).}
    \item We \omcr{1}{develop} and analyze four potential ways to mitigate \ieycr{0}{\X{}}. We adapt and evaluate the effectiveness of the industry's state-of-the-art RowHammer mitigation, PRAC, and show that \omcr{0}{adapted} PRAC \omcr{0}{incurs large} performance overheads \omcr{0}{to mitigate \X{}}. 
    \ieycr{1}{\item Our takeaway lessons (from both characterization and mitigation) call for future work on understanding the underlying device-level causes of \X{} bitflips and other innovative solutions to mitigate \X{} bitflips to enable read-disturbance-resilient future \gls{pud} systems.}
\end{itemize}

\glsresetall
\section{Background}

\subsection{Dynamic Random Access Memory (DRAM)}
\label{sec:dram_organization}

\begin{figure}[ht]
    \centering
    \includegraphics[width=\linewidth]{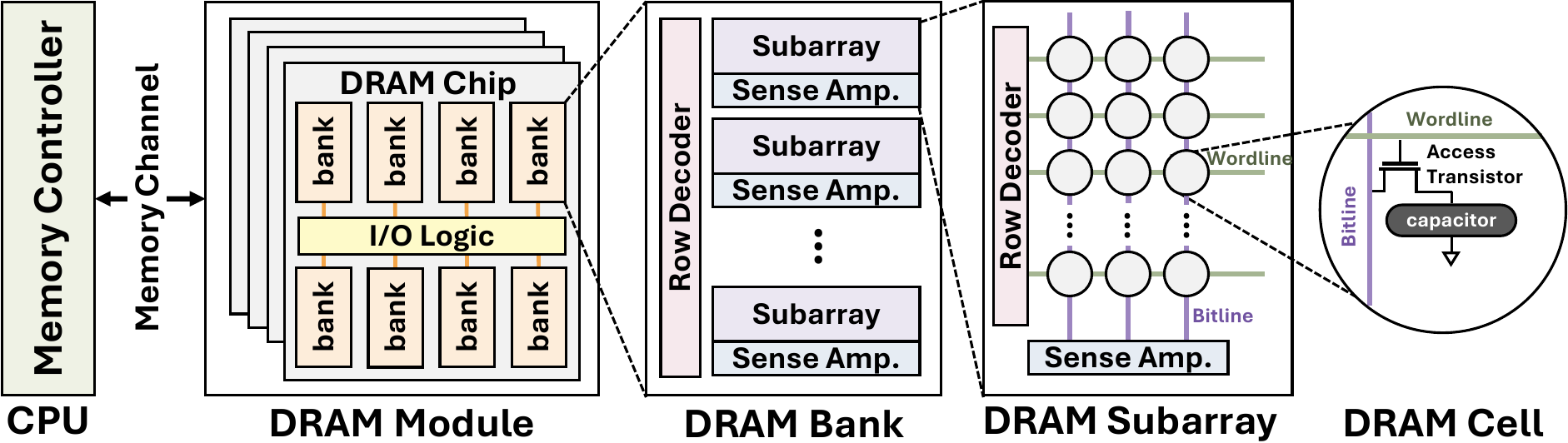}
    \caption{Hierarchical organization of modern DRAM.}
    \label{fig:dram_organization}
\end{figure}

\noindent\textbf{DRAM Organization.}~\figref{fig:dram_organization} shows the hierarchical organization of a modern DRAM-based main memory. The memory controller connects to a DRAM module over a memory channel. A module contains one or multiple DRAM ranks that time-share the memory channel. A rank consists of multiple DRAM chips.
Each DRAM chip has multiple DRAM banks, each containing multiple subarrays. 

Within a subarray, DRAM cells form a two-dimensional structure interconnected over \textit{bitlines} and \textit{wordlines}. The row decoder in a subarray decodes the row address and drives one wordline out of many. A row of DRAM cells on the same wordline is referred to as a DRAM \emph{row}. The DRAM cells in the same \emph{column} are connected to the sense amplifier via a bitline.
A DRAM cell stores \omcr{0}{a} binary data value in the form of electrical charge on a capacitor (\vdd{} or 0~V), and this data is accessed through an access transistor, driven by the wordline to connect the cell capacitor to the bitline.

\noindent\textbf{DRAM Access.}
The memorycontroller serves memory access requests by issuing DRAM
commands, e.g., row activation (\act{}), bank precharge (\pre{}), data read
(\rd{}), data write (\wri{}), and refresh ($REF$) while respecting certain timing
parameters to guarantee correct operation~\cite{jedec2020lpddr5,
jedec2015lpddr4,jedecddr,jedec2020ddr4,jedec2012ddr3,jedecddr5c,jedec2021hbm,lee2013tiered,kim2012case,lee2015adaptive,mutlu2008parallelism,kim2010atlas}.
The memory controller issues an \act{} command alongside the bank
address and row address corresponding to the memory request's address
to activate a DRAM row. During the row activation process, a
DRAM cell \omcr{0}{shares} its charge, and thus, its initial charge \omcr{0}{level} needs to be restored
(via a process called \emph{charge restoration}). The latency from the start of
a row activation until the completion of the DRAM cell's charge restoration is
called \emph{$t_{RAS}$}. To access another row in an already activated DRAM
bank, the memory controller must issue a \pre{} command to close the opened row
and prepare the bank for a new activation. The minimum latency between issuing a \pre{} command and opening a row with an \act{} command is called \emph{$t_{RP}$}.

\noindent\textbf{DRAM Refresh.}
A DRAM cell is inherently leaky and thus loses its stored electrical
charge over time\ieycr{0}{~\cite{liu2012raidr,qureshi2015avatar,liu2013experimental}}. To maintain data integrity, a DRAM cell is periodically
refreshed with a {time interval called the \emph{$t_{REFW}$}, which is
typically} \SI{64}{\milli\second} (e.g.,\ieycr{0}{~\cite{jedec2012ddr3, jedec2020ddr4,
micron2014ddr4,liu2013experimental}}) at normal operating temperature (i.e., up to
\SI{85}{\celsius}).  
To refresh all cells in a timely manner, the memory controller
periodically issues a refresh (\texttt{REF}) command with a time interval called
the \emph{$t_{REFI}$}, which is typically \SI{7.8}{\micro\second}
(e.g.,\ieycr{0}{~\cite{jedec2012ddr3, jedec2020ddr4, micron2014ddr4, hassan2021utrr}}) or
\SI{3.9}{\micro\second} (e.g.,~\cite{jedec2015lpddr4, jedecddr5c,
jedec2020lpddr5}) at normal operating temperature.

\subsection{DRAM Read Disturbance}
\label{subsec:back_disturbance}
Read disturbance is the phenomenon that reading data from a memory or storage device causes physical disturbance (e.g., voltage deviation, electron injection, electron trapping) on another piece of data that is \emph{not} accessed but physically located nearby the accessed data. Two prime examples of read disturbance in modern DRAM chips\omcr{0}{~\cite{mutlu2019rowhammer,mutlu2023fundamentally,mutlu2017rowhammer,mutlu2023retrospective}} are RowHammer~\cite{kim2014flipping}, and RowPress~\cite{luo2023rowpress}, where repeatedly accessing (hammering) or
keeping active (pressing) a DRAM row induces bitflips in physically nearby DRAM
rows. In RowHammer and RowPress terminology, the row that is
hammered or pressed is called the \emph{aggressor} row, and the row that
experiences bitflips the \emph{victim} row. For read disturbance bitflips to occur, 1)~the aggressor row needs to be activated more than a certain threshold value, defined as \gls{hcfirst}~\cite{kim2020revisiting} and/or 2) the aggressor row needs to be open for a long period of time  (i.e., $\DRAMTIMING{AggON} > \DRAMTIMING{RAS}$){~\cite{luo2023rowpress}}.

\subsection{Processing-using-DRAM (PuD)}
\label{subsec:back_pud}
\gls{pud} is an emerging paradigm that can alleviate the bottleneck caused by frequent data movement between processing elements (e.g., CPU) and main memory~\pudAllCitations{}. PuD enables massively parallel in-DRAM computation by leveraging intrinsic analog operational properties of the DRAM circuitry.
\omcr{0}{\ieycr{1}{M}any \gls{pud} works~\cite{gao2019computedram,olgun2023dram,gao2022frac,yuksel2024simultaneous,olgun2022pidram,yuksel2024functionally, seshadri2013rowclone,seshadri2016buddy,seshadri2015fast,seshadri2016processing,seshadri2017ambit,seshadri2017simple,seshadri2018rowclone,seshadri2019dram,mutlu2024memory,mutlu2025memory} \ieycr{1}{enables}} \one{} in-DRAM data copy \& initialization \ieycr{1}{by leveraging} \gls{comraL} and \gls{simraL} and \two{} in-DRAM bitwise operations by leveraging \acrshort{simra}.

\noindent\textbf{In-DRAM Data Copy \& Initialization.}
RowClone~\cite{seshadri2013rowclone} enables data movement within a subarray at a row granularity by modifying DRAM circuitry. RowClone alleviates the energy and execution time costs of transferring data between the DRAM and the processing units. Prior works~\cite{gao2019computedram,olgun2022pidram} experimentally demonstrate that the RowClone operation can be performed in COTS DRAM chips by enabling \omcr{0}{\emph{consecutive activation of two rows}} in the same subarray\ieycr{0}{, which we call CoMRA}. A recent work~\cite{yuksel2024simultaneous} demonstrates that \gls{cots} DRAM chips can copy one source row to up to 31 different destination rows by simultaneously activating up to 32 rows in the same subarray.

\noindent\textbf{In-DRAM Bitwise Operations.}
Prior works~\cite{seshadri2017ambit,seshadri2015fast} demonstrates that 1) \omcr{0}{simultaneously} activating three DRAM rows leads to the computation of the bitwise MAJority function (and thus the AND and OR functions) on the contents of the three rows due to the charge sharing principles that govern the operation of the shared bitlines and sense amplifiers, 
2)~bitwise NOT of a row can be performed through the sense amplifier, with modifications to DRAM circuitry.
Many operations envisioned by the\omcr{0}{se} work\omcr{0}{s~\cite{seshadri2017ambit,seshadri2015fast}} can \emph{already} be performed in \emph{real unmodified} \gls{cots} DRAM chips, by violating manufacturer-recommended DRAM timing parameters\omcr{0}{~\cite{gao2019computedram,olgun2023dram,gao2022frac,yuksel2024simultaneous,yuksel2024functionally}}.
Recent works show that COTS DRAM chips can perform 1) the bitwise MAJ operation with up to nine inputs (i.e., MAJ3, MAJ5, MAJ7, and MAJ9) by simultaneously activating multiple rows in the same subarray~\cite{gao2019computedram,olgun2023dram,gao2022frac,yuksel2024simultaneous} and 
2) up to 16-input AND, NAND, OR, NOR operations, and NOT operation by simultaneously activating multiple rows in two neighboring subarrays~\cite{yuksel2024functionally}. 

\subsection{\ieycr{1}{Motivation}}
\ieycr{1}{\gls{pud} is a promising paradigm that has the potential to reduce or eliminate costly data movement between processing elements and main memory~\pudAllCitations{}. Many \gls{pud} techniques (\secref{subsec:back_pud}) leverage an analog DRAM operation called multiple-row activation, where multiple DRAM rows are activated simultaneously or in quick succession to perform in-DRAM processing~\cite{deoliveira2024mimdram,oliveira2025proteus,hajinazar2021simdram,gao2019computedram,gao2022frac,yuksel2024functionally,yuksel2024simultaneous,mutlu2024memory,olgun2021quac, olgun2022pidram,olgun2023dram,yaglikci2022hira,seshadri2013rowclone,seshadri2015fast,seshadri2016buddy,seshadri2016processing,seshadri2017ambit,seshadri2017simple,seshadri2018rowclone,seshadri2019dram}. Multiple-row activation fundamentally differs from standard DRAM accesses, where a single row is activated at a time (\secref{sec:dram_organization}). This fundamental difference could result in significant implications for future \gls{pud}-enabled systems, as repeatedly activating \emph{even} a single DRAM row can induce bitflips in other unaccessed DRAM rows due to DRAM read disturbance phenomena (\secref{subsec:back_disturbance}). Unfortunately, no prior work explores the read disturbance effects of multiple-row activation. Our goal in this paper is to close this gap. We aim to empirically understand, characterize, and provide insights into the interaction between read disturbance and multiple-row activation.}
\section{Metholodogy}
We describe our \gls{cots} DRAM \omcr{0}{chip} testing infrastructure (\secref{subsec:infra}) and the \gls{cots} DDR4 chips
tested for our characterization study (\secref{subsec:tested_chips}). We explain the methodology of our different characterization experiments in
their corresponding sections, \secref{sec:comra}, \secref{sec:simra}, and \secref{sec:combined}.

\subsection{COTS DRAM Testing Infrastructure}
\label{subsec:infra}
\begin{figure}[ht]
\centering
\includegraphics[width=\linewidth]{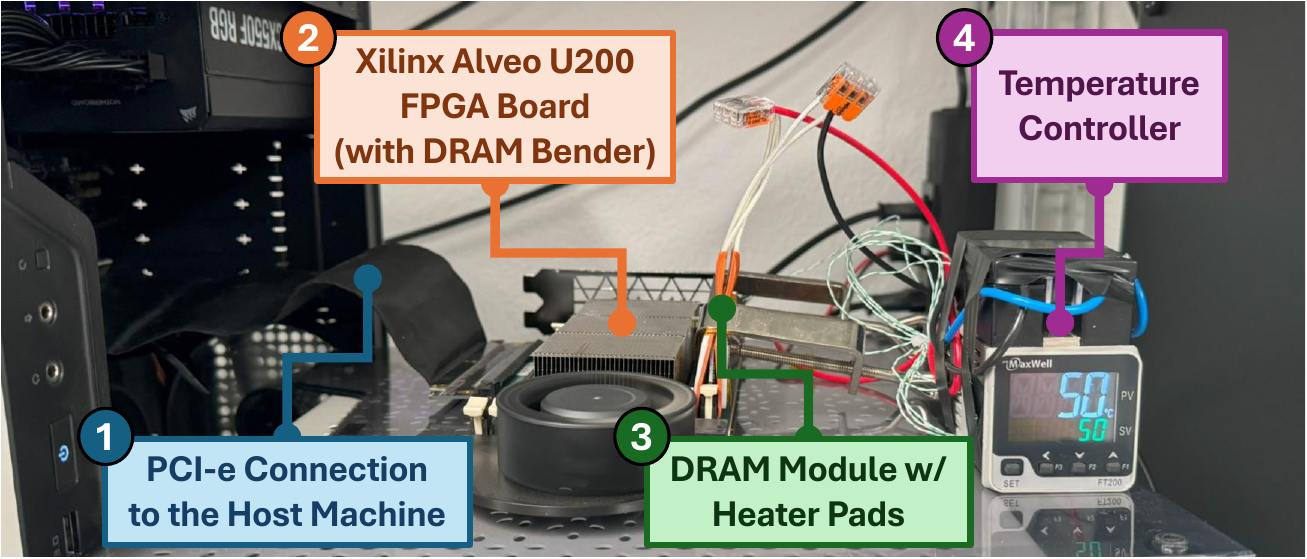}
\caption{Our DRAM Bender~\cite{olgun2023dram} based experimental setup.}
\label{fig:infra}
\end{figure}

We conduct COTS DRAM chip experiments using DRAM Bender~\cite{safari-drambender, olgun2023dram} \omcr{0}{(built upon SoftMC~\cite{hassan2017softmc,softmcgithub})}, an FPGA-based DDR4 testing infrastructure that provides precise control of DDR4 commands. \figref{fig:infra} shows our experimental setup that consists of four main components: 1) a host machine that generates the test program and collects results, 2) an FPGA development board~\cite{alveo_u200}, programmed with DRAM Bender, 3) a thermocouple temperature sensor and heater pads pressed against the DRAM chips to maintain a target temperature level, and 4) a temperature controller~\cite{maxwellFT200} that keeps the temperature at the desired level. 

\noindent\textbf{Eliminating Interference Sources.}
To observe read disturbance bitflips \omcr{0}{at} the circuit level,
we eliminate potential sources of interference, by taking \param{three} measures, similar to the methodology used by prior works~\cite{kim2020revisiting, orosa2021deeper, yaglikci2022understanding, hassan2021utrr, luo2023rowpress}.
First, we disable periodic refresh during the execution of test programs to prevent potential on-DRAM-die TRR mechanisms~\cite{frigo2020trrespass, hassan2021utrr} from refreshing victim rows so that we can observe the DRAM chip's behavior at the circuit-level.
Second, we strictly bound the execution time of test programs within the refresh window of the DRAM chips to avoid data retention failures interfering with read disturbance failures.
Third, we {verify} that the modules and chips have neither rank-level nor on-die ECC~\cite{patel2020beer, patel2021harp}.
{With these measures,} we directly observe and analyze all bitflips without interference.

\subsection{COTS DDR4 DRAM Chips Tested}
\label{subsec:tested_chips}

Table~\ref{tab:dram_chips} provides the \nCHIPS{} (\nMODULES{}) COTS DDR4 DRAM chips (modules) along with the chip manufacturer (Chip Mfr.), number of modules (\#Modules), number of chips (\#Chips), die revision (Die Rev.), chip density (Density), and chip organization (Org.).

\begin{scriptsize}
\begin{table}[ht]
\centering
\caption{Summary of DDR4 DRAM chips tested.}
\label{tab:dram_chips}
\resizebox{\columnwidth}{!}{%
\renewcommand{\arraystretch}{0.87}
\begin{tabular}{cccccc}
\textbf{Chip Mfr.} & \textbf{\#Modules} & \textbf{\#Chips} &  \textbf{Die Rev.} & \textbf{Density} & \textbf{Org.}  \\

\hline\hline
& 1 & \param{8} & A & 4Gb  & x8 \\
SK Hynix & 8 & \param{64} & A & 8Gb  & x8  \\ 
 & 2 & \param{16} & C & 16Gb  & x8   \\
& 6 & \param{48} & D & 8Gb  & x8 \\

\midrule

& 1 & \param{8} & B  & 4Gb & x8 \\
Micron  & 4 & \param{32} & E  & 16Gb & x16  \\
& 4 & \param{32} & F  & 16Gb & x8 \\
 & 2 & \param{16} & R  & 8Gb & x8 \\

\midrule
 & 1 & \param{8} & A  & 16Gb & x8 \\
& 5 & \param{40} & B  & 16Gb & x8 \\
Samsung  & 1 & \param{4} & C  & 4Gb & x16 \\
& 1 & \param{8} & C  & 16Gb & x8  \\
& 1 & \param{8} & E  & 4Gb & x8 \\

 \midrule

Nanya   & 3 & \param{24} & C  & 8Gb & x8 \\
\bottomrule
\end{tabular}
}
\end{table}
\end{scriptsize}

\noindent\textbf{Logical-to-Physical Row Mapping.}
DRAM manufacturers use mapping schemes to translate logical (memory-controller-visible) addresses to physical row addresses~\cite{kim2014flipping, smith1981laser, horiguchi1997redundancy, keeth2001dram, itoh2013vlsi, liu2013experimental,seshadri2015gather, khan2016parbor, khan2017detecting, lee2017design, tatar2018defeating, barenghi2018software, cojocar2020rowhammer,  patel2020beer,kim2012case,frigo2020trrespass}. To account for in-DRAM row address mapping, we reverse engineer the physical row address layout \omcr{0}{in all chips}, following the prior works' methodology~\cite{kim2020revisiting, orosa2021deeper, yaglikci2022understanding, luo2023rowpress}.

\section{Read Disturbance Effect of Consecutive Multiple-Row Activation (CoMRA) in\\COTS DRAM Chips}
\label{sec:comra}
We demonstrate the read disturbance effect of consecutive multiple row activation \comra{} in \gls{cots} DRAM chips. \comra{} is used to perform in-DRAM data copy operations in real DRAM chips~\cite{gao2019computedram,olgun2021quac,yuksel2024functionally,yuksel2024simultaneous,gao2022frac,olgun2022pidram,olgun2023dram}. We repeatedly perform consecutive activation of two rows (source row and destination row) to copy \omcr{0}{a} source row's content into \omcr{0}{a} destination row. This section describes our key idea to \ous{hammer with} \comra{} (\secref{subsec:key_comra}), our experimental methodology for understanding the read disturbance vulnerability caused by the \comra{} operation (\secref{subsec:method_comra}), and presents our COTS DRAM chip characterization results (\secref{subsec:char_comra}).

\subsection{Hammering with CoMRA}
\label{subsec:key_comra}
\begin{figure}[ht]
    \centering
    \includegraphics[width=\linewidth]{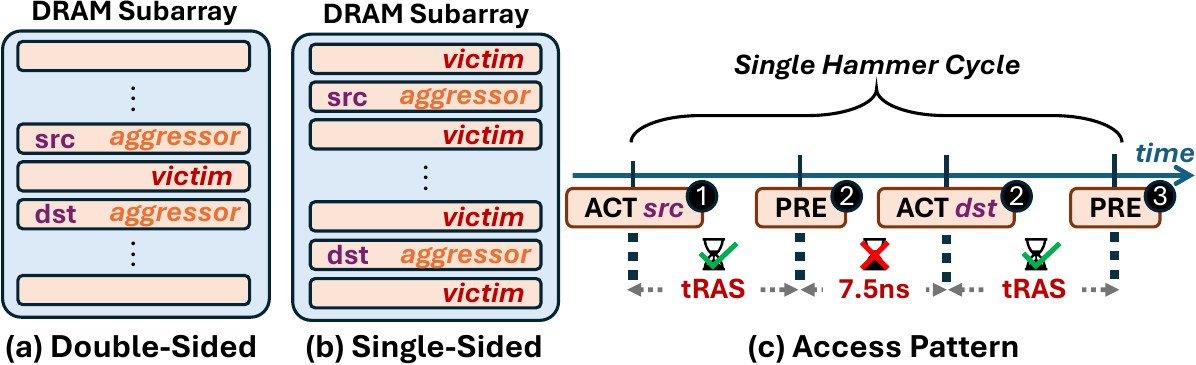}
    \caption{Example of a double-sided \comra{} attack (a), a single-sided \comra{} attack (b), and the access pattern of double-sided and single-sided \comra{} (c).}
    \label{fig:comra_mech}
\end{figure}

\noindent\textbf{Key Idea.}
We exploit the key property of \comra{} to induce read disturbance bitflips in COTS DRAM chips: activating two rows consecutively in quick succession to \ous{repeatedly} perform in-DRAM data copy operations.
To do so, we \ous{repeatedly} perform consecutive activation of \omcr{0}{a} source row (\src{}) and \omcr{0}{a} destination row (\dst{}) in the same subarray to induce read disturbance bitflips in neighboring rows of the \src{} and \dst{}.
In this scenario, \src{} and \dst{} are the aggressor rows, and their neighboring rows are victim rows. 
\figref{fig:comra_mech} illustrates our key idea to hammer with \comra{}.
Depending on the location of source and destination rows, we can \ous{either craft} \one{}~a double-sided attack where \src{} and \dst{} are sandwiching a victim row (\figref{fig:comra_mech}a) \ous{or} \two{}~a single-sided attack where \src{} and \dst{} are far away (\figref{fig:comra_mech}b).

\noindent\textbf{Access Pattern \& Operation.}
\figref{fig:comra_mech}c shows \omcr{0}{both single-sided and double-sided access patterns} of \ous{hammering with \comra{}}. Our attack consists of three key steps.
First, we issue \act{} \omcr{1}{(ACTIVATE)} \src{} to activate the \src{} row (\dingOne{}) and wait for $\tras{}$, which ensures the \ous{sense amplifier senses the} data in the \src{} row and \ous{restores the charges of cells}.
Second, we issue \pre{} \omcr{1}{(PRECHARGE)} and \act{}~\dst{} back-to-back by violating the $\trp{}$ timing parameter to activate the \dst{} while bitlines and sense amplifiers still have \src{}'s content~\cite{yuksel2024simultaneous,gao2019computedram,olgun2022pidram} (\dingTwo{}).
Doing so copies the \src{} row's data to the \dst{} row~\cite{yuksel2024simultaneous,olgun2021quac,yuksel2024functionally,yuksel2024simultaneous,yaglikci2024svard,olgun2022pidram,olgun2023dram}.
Third, we \ous{wait for $\tras{}$ to ensure \omcr{0}{that} the in-DRAM copy operation completes and issue a} \pre{} \ieycr{0}{to prepare the bank for the next attack access} (\dingThree{}).
We call this three-step in-DRAM copy procedure one hammer cycle. To induce read disturbance bitflips in the victim rows, we repeatedly perform \comra{}.

\subsection{Experimental Methodology}
\label{subsec:method_comra}
\noindent\textbf{DRAM Subarray Boundaries.} Understanding the read disturbance effect of \omcr{0}{the} in-DRAM copy operation in \gls{cots} DRAM chips requires reverse engineering DRAM subarray boundaries \iey{as \omcr{0}{the} in-DRAM copy operation works \omcr{0}{only} if the source and destination \omcr{0}{rows are} locat\omcr{0}{ed} in the same subarray~\cite{seshadri2013rowclone,seshadri2018rowclone,gao2019computedram,olgun2022pidram,olgun2021quac,yuksel2024functionally,yuksel2024simultaneous,yaglikci2024svard}}. 
We repeatedly perform the RowClone operation for \emph{every} possible source and destination row address in each tested bank. When we observe that the destination row gets the same content as the source row after the in-DRAM data copy, we conclude that the source row and the destination row are in the same subarray. Based on this observation, we reverse engineer the subarray boundaries and determine which rows are in the same subarray.

\noindent\textbf{Read Disturbance Vulnerability Metric.}
To characterize a DRAM module's vulnerability to read disturbance, we examine~\glsentrylong{hcfirst}~(\gls{hcfirst}), where we count the hammer cycles (i.e., each pair of activations to the \src{} and \dst{} rows as one hammer). A \omcr{0}{lower} \gls{hcfirst} indicates a \omcr{0}{higher} vulnerability to read disturbance.

\noindent\textbf{\pmb{\hcfirst{}} Algorithm.}
For every tested parameter we evaluate (e.g., data pattern and temperature), we find the \gls{hcfirst} for each tested victim row using the bisection-method algorithm used by prior works~\cite{orosa2021deeper,yaglikci2022understanding,luo2023rowpress}. We terminate the \gls{hcfirst} search when the difference between the current and previous \gls{hcfirst} measurements is no larger than 1\% of the previous measurements. For every tested row, we repeat the \gls{hcfirst} search five times and report the minimum \gls{hcfirst} value we observe.

\noindent\textbf{Victim Row Location in the Subarray.}
To understand the effects of spatial variation on read disturbance, we analyze how the location of a victim row in a subarray affects the read disturbance vulnerability. We categorize a victim row's location in a subarray into five regions: \one{} "Beginning": the first \omcr{0}{20\%} rows in the subarray (e.g., the first 100 rows in a subarray with 500 rows, row 0 to row 99), \two{} "Beginning-Middle": the second \omcr{0}{20\%} rows in the subarray (e.g., row 100 to row 199), "Middle": the third \omcr{0}{20\%} rows in the subarray (e.g., row 200 to row 299), "Middle-End": the fourth \omcr{0}{20\%} rows in the subarray (e.g., row 300 to row 399), and  \five{} "End": the last \omcr{0}{20\%} rows in the subarray (e.g., row 400 to row 499). 

\noindent\textbf{Data Pattern.}
 We use the four data patterns (\texttt{0x00},~\texttt{0xFF},~\texttt{0xAA}, and~\texttt{0x55}) that are widely used in memory reliability testing~\cite{vandegoor2002address,khan2014efficacy} and by prior work on DRAM characterization (e.g.,~\cite{kim2014flipping,kim2020revisiting,orosa2021deeper,luo2023rowpress,yaglikci2024svard,olgun2025variable,luo2024experimental,tugrul2025understanding,olgun2024read}). We fill aggressor rows (\src{} and \dst{}) with these data patterns while initializing victim rows with the negated data pattern (e.g., if aggressor rows are~\texttt{0x00}, victim rows are~\texttt{0xFF}). For each DRAM row, we define the worst-case data pattern (\emph{WCDP}) as the data pattern that causes the lowest \gls{hcfirst}. All experiments are conducted using WCDP unless stated otherwise.

\noindent\textbf{Temperature.} 
We perform our experiments at four temperature levels: 50$^{\circ}$C, 60$^{\circ}$C, 70$^{\circ}$C, and 80$^{\circ}$C. All experiments are conducted at 80$^{\circ}$C unless stated otherwise.

\noindent\textbf{Timing Delay.}
We sweep the timing delay between \pre{}$\rightarrow$\act{}~\dst{} command presented in \figref{fig:comra_mech}c. All experiments are conducted \ous{with a timing delay of \omcr{0}{violated} $7.5ns$ (as in \figref{fig:comra_mech}c)} unless stated otherwise.

\noindent\textbf{Number of Instances Tested.} \ieycr{0}{To maintain a reasonable testing time, w}e select six subarrays in a bank per DRAM module: two subarrays from the beginning of the bank, two subarrays from the middle of the bank, and two subarrays from the end of the bank. Within each subarray, we test all rows.

\subsection{COTS DRAM Chip Characterization}
\label{subsec:char_comra}
This section presents our characterization of the read disturbance caused by consecutive activation of two rows in \gls{cots} DRAM chips.

\begin{figure}[ht]
    \centering
    \includegraphics[width=\linewidth]{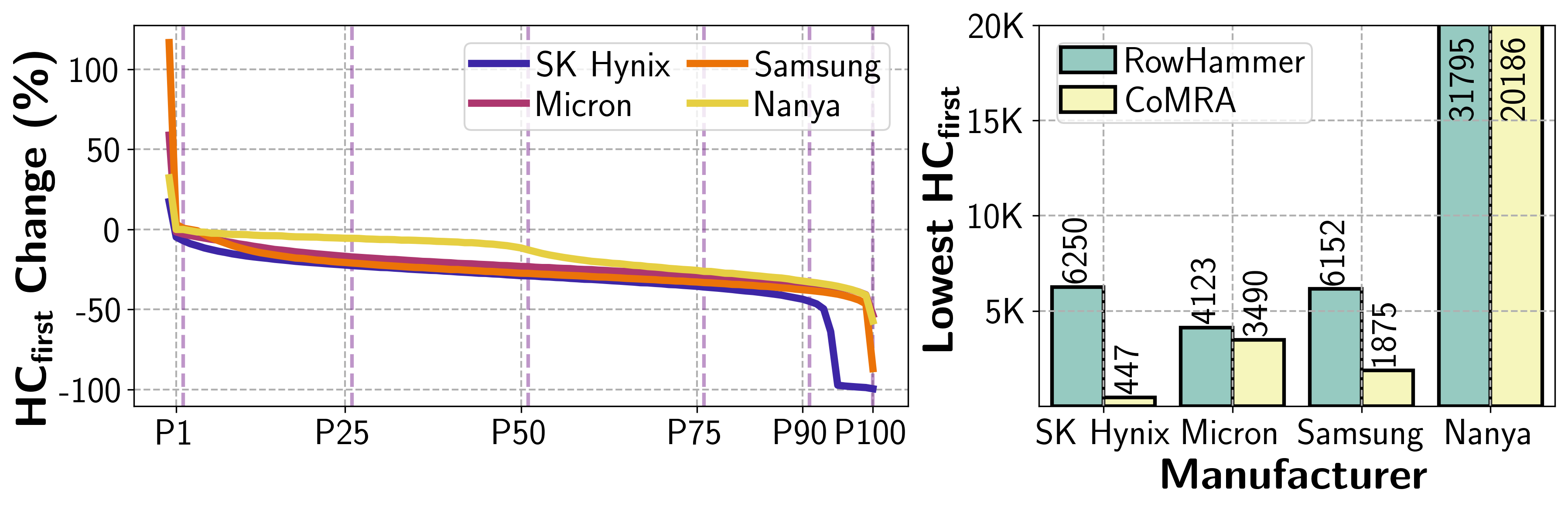}
    \caption{Distribution of the change \omcr{1}{in} \gls{hcfirst} \omcr{1}{with} double-sided \comra{} compared to double-sided RowHammer (left) and the lowest \gls{hcfirst} observed \omcr{1}{with} double-sided \comra{} and RowHammer \omcr{1}{(right)} for each manufacturer.}
    \label{fig:rc_rh_ds}
\end{figure}

\noindent\textbf{Double-Sided \comra{} vs. RowHammer.}
We investigate the variation in \gls{hcfirst} across rows when we perform double-sided \comra{} and double-sided RowHammer. In \figref{fig:rc_rh_ds}, the left plot shows the distribution of the change in \gls{hcfirst} (in percentage) when we perform double-sided \comra{} \ieycr{0}{compared to} double-sided RowHammer for \ieycr{0}{all tested \omcr{1}{chips from} \param{four}} manufacturers. 
The x-axis represents the percentage of all \ieycr{0}{victim} rows \ieycr{1}{from all tested chips from all tested manufacturers}, sorted from the most positive \gls{hcfirst} change \ous{(i.e., \comra{} has higher \gls{hcfirst} compared to RowHammer)} to the most negative \gls{hcfirst} change \ous{(i.e., \comra{} has lower \gls{hcfirst} compared to RowHammer)}.
In \figref{fig:rc_rh_ds}, the right plot shows the lowest \gls{hcfirst} observed across \ieycr{1}{all tested} \nCHIPS{} DRAM chips \ieycr{1}{from four manufacturers} for double-sided \comra{} and RowHammer.

\observation{\ous{Hammering with double-sided} \comra{} significantly decreases \gls{hcfirst} \ieycr{0}{compared to double-sided RowHammer}.}
We observe that \ous{when rows are hammered with double-sided \comra{}}, the lowest \gls{hcfirst} observed is 447, 3490, 1875, and 20186 for SK Hynix, Micron, Samsung, and Nanya chips, respectively.
Compared to double-sided RowHammer, double-sided \comra{} decreases the lowest \gls{hcfirst} by 13.98$\times$, 1.18$\times$, 3.28$\times$, and 1.58$\times$ for SK Hynix, Micron, Samsung, and Nanya chips, respectively.

\hluo{We hypothesize that the reason for \ous{double-sided} \comra{}'s lower \gls{hcfirst} compared to double-sided RowHammer is that the reduced interval between closing of the \src{} wordline and the activation of the \dst{} wordline enhances trap-assisted electron migration from near the neighboring aggressor wordline towards the victim node~\cite{yang2019trap, zhou2023double} (i.e., trapped electron density is higher when the neighboring aggressor wordline is \emph{just} closed).} We call for future research to fundamentally understand \comra{}'s read disturbance.

\observation{\ous{Hammering with double-sided} \comra{} decreases \gls{hcfirst} for a large fraction of DRAM rows.}
\ieycr{1}{In the left plot of \figref{fig:rc_rh_ds}, we observe that} compared to double-sided RowHammer, 99\% of DRAM rows experience \ieycr{0}{the first} bitflip with fewer activation counts when performing double-sided \comra{} for all four manufacturers. 

\takeaway{\comra{} exacerbates DRAM's vulnerability to read disturbance in all four major manufacturers.}

\begin{figure}[ht]
    \centering
    \includegraphics[width=\linewidth]{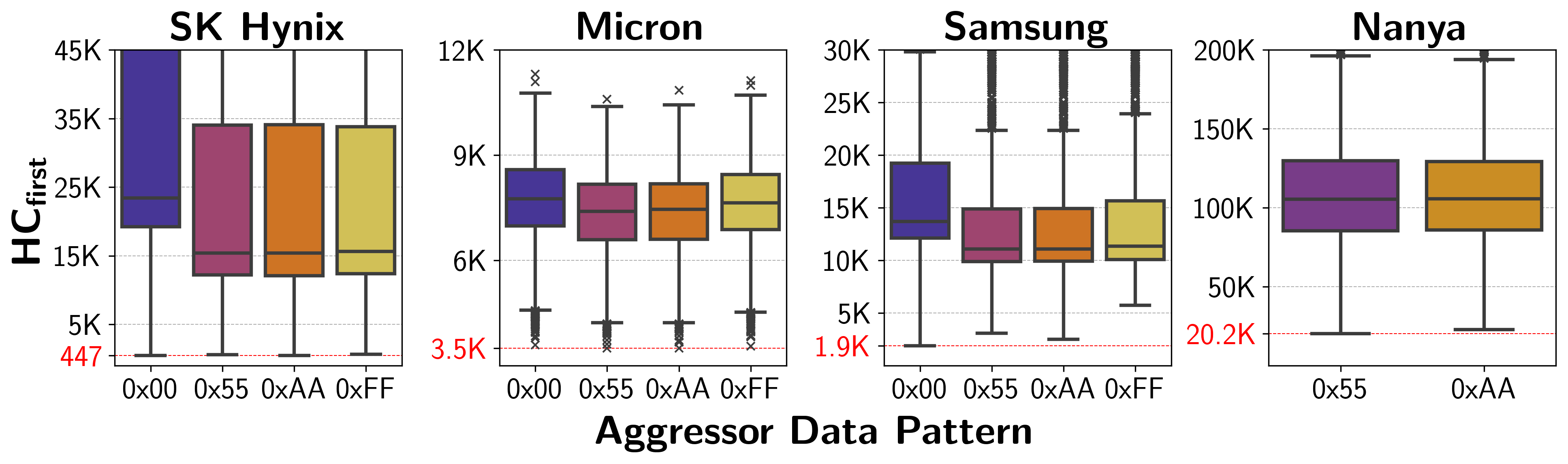}
    \caption{\gls{hcfirst} distribution of double-sided \comra{} with different \ieycr{1}{aggressor} data patterns. \omcr{1}{Victim rows have negated aggressor data pattern}.}
    \label{fig:rc_dp}
\end{figure}

\noindent\textbf{Data Pattern.}
We analyze \ous{the effect of data pattern on} \gls{hcfirst} across DRAM rows when we perform double-sided \comra{}.
\figref{fig:rc_dp} shows the \gls{hcfirst} distribution across all tested DRAM rows (y-axis) for four data patterns (x-axis).\footnote{Due to \ieycr{0}{the} complicated true/anti cell pattern of Nanya chips, we could not observe bitflips within a refresh window with 0xFF and 0x00 data patterns.} 

\observation{Che{c}kerBoard pattern (i.e., 0x55/0xAA) is, in general, the most effective data pattern among the ones tested.}
\ous{When performing double-sided \comra{}}, in most cases, 0x55/0xAA is the most effective at inducing bitflips compared to 0x00/0xFF, similar to RowHammer~\cite{orosa2021deeper,luo2023rowpress, kim2014flipping}. For example, in Samsung \omcr{1}{chips}, average \gls{hcfirst} for 0x55 is \param{17346}, whereas for 0x00 is \param{21423}. However, \ous{in some cases}, we also observe that the worst-case data pattern is not the CheckerBoard pattern, similar to prior works~\cite{orosa2021deeper, kim2020revisiting, luo2023rowpress, olgun2024read, kim2014flipping}.

\begin{figure}[ht]
    \centering
    \includegraphics[width=\linewidth]{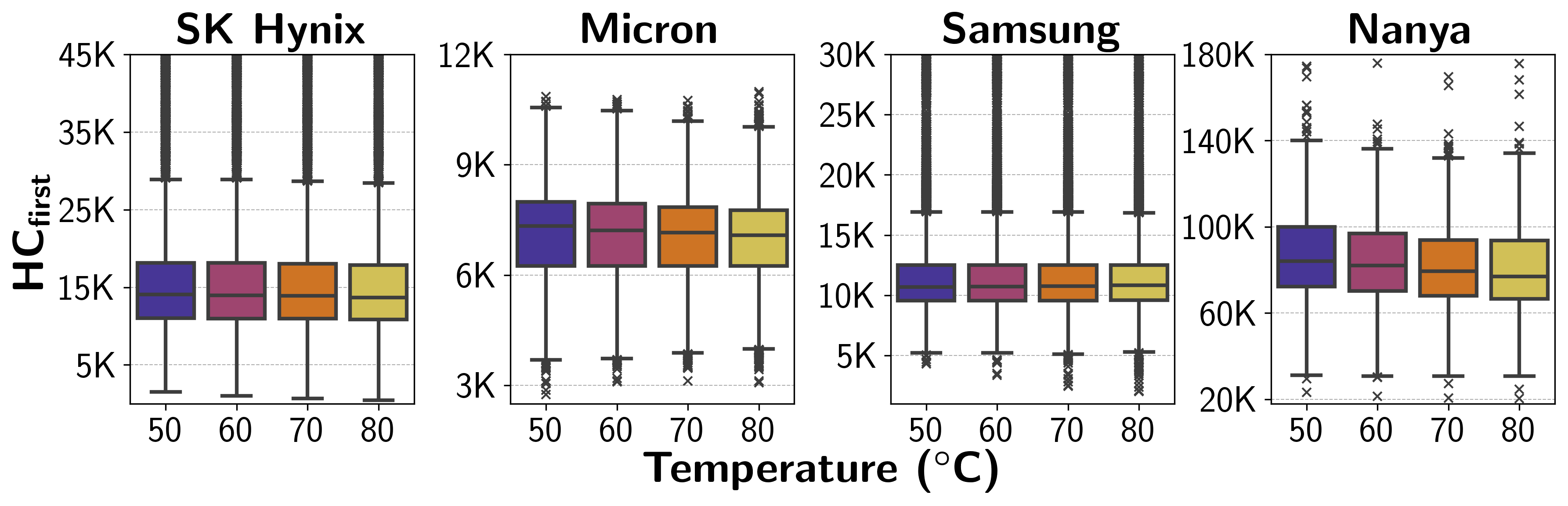}
    \caption{\gls{hcfirst} distribution of \ous{hammering with} double-sided \comra{} at different DRAM chip temperatures.}
    \label{fig:rc_temp}
\end{figure}

\noindent\textbf{Temperature.}
\figref{fig:rc_temp} shows the \gls{hcfirst} distribution of \ous{hammering with} double-sided \comra{} at four different temperature levels: 50$^{\circ}$C, 60$^{\circ}$C, 70$^{\circ}$C, and 80$^{\circ}$C. Each subplot is dedicated to a different manufacturer, where the x-axis shows the tested temperature levels.

\observation{\ous{Read disturbance effects of hammering with double-sided} \comra{} tend to get worse as temperature increases.}
We observe that {when we perform double-sided \comra{}}, \ous{as temperature increases from 50$^{\circ}$C to 80$^{\circ}$C, the lowest \gls{hcfirst}} decreases by \param{3.45$\times$}, \param{2.13$\times$}, and \param{1.14$\times$} for SK Hynix, Samsung, and Nanya, \ous{respectively}.
\ous{On the other hand, for Micron}, the effect is the opposite, \ous{where the lowest \gls{hcfirst} increases as the temperature increases} (e.g., by \param{1.14$\times$} from 50$^{\circ}$C to 80$^{\circ}$C). A prior work~\cite{orosa2021deeper} hypothesizes that the relation between read disturbance vulnerability and temperature is caused by the nonmonotonic behavior of charge-trapping characteristics of DRAM cells, which results in individual DRAM rows exhibiting different behavior. This hypothesis could also explain Micron's \gls{hcfirst} trend as we sweep temperature.

\takeaway{Hammering with \comra{} is affected by temperature and data pattern. Worst-case data pattern and temperature tend to differ for individual DRAM rows.}

\begin{figure}[ht]
    \centering
    \includegraphics[width=\linewidth]{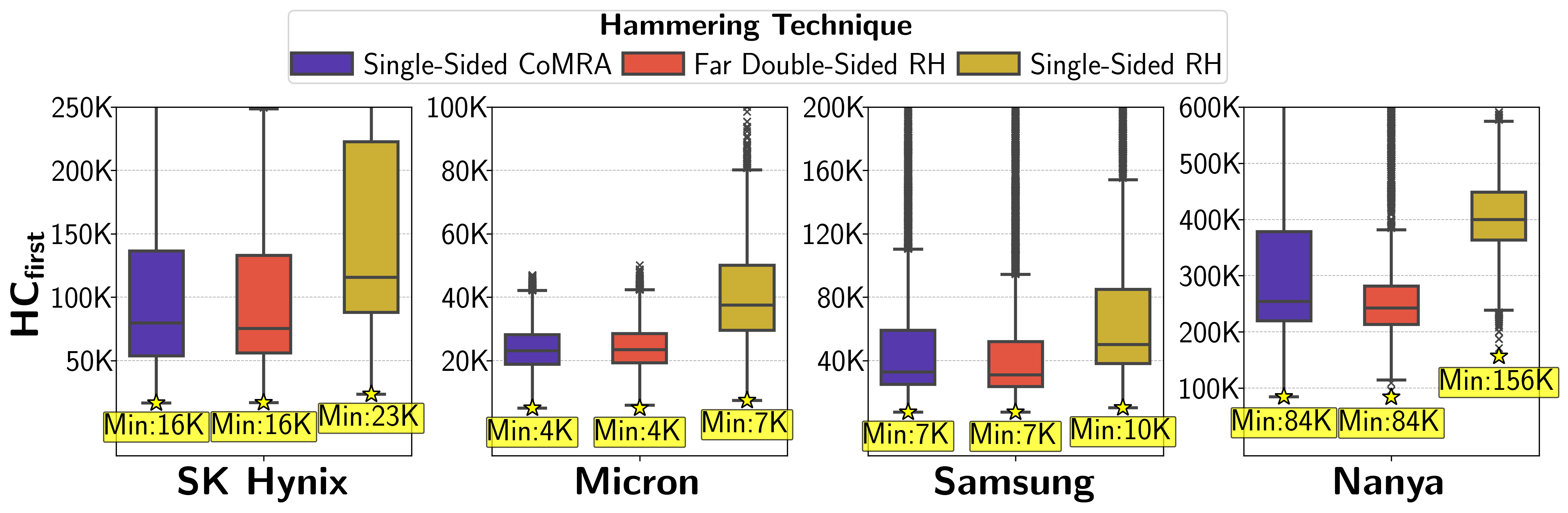}
    \caption{\gls{hcfirst} of single-sided \comra{} and RowHammer.}
    \label{fig:rc_rh_ss}
\end{figure}

\noindent\textbf{Single-Sided \comra{} vs. RowHammer.}
We analyze the \gls{hcfirst} variation across DRAM rows for three techniques:
\one{} single-sided \comra{} where \src{} and \dst{} are far away from each other (e.g., 100 rows apart),
\two{} single-sided RowHammer where we keep hammering one aggressor row, and
\three{} far double-sided RowHammer where the access pattern is the same as \ous{single-sided \comra{}} except we use $\trp{}$ latency between \pre{} to \act{}~\dst{}.
We evaluate the far double-sided RowHammer \ous{to observe the effect of reduced \pre{} to \act{}~\dst{} latency in} single-sided \comra{}.
\figref{fig:rc_rh_ss} shows the \gls{hcfirst} distribution of these three techniques, \ous{where each} subplot is dedicated to a different manufacturer \ous{and colored boxes present different techniques}.
We highlight the lowest \ous{observed \gls{hcfirst} of} each technique for every manufacturer with a yellow rectangle.

\observation{Single-sided \comra{} decreases \gls{hcfirst} compared to single-sided RowHammer and \ous{exhibits} similar \gls{hcfirst} distribution with far double-sided RowHammer.}
We observe that, for all four manufacturers, single-sided \comra{} \one{} is more effective than single-sided RowHammer, and \two{} performs similarly \ous{to} far double-sided RowHammer.
For example, in SK Hynix \omcr{1}{chips}, the lowest \ous{observed} \gls{hcfirst} for single-sided \comra{} is \param{16495}, which is \param{1.42$\times$} and \param{1.02$\times$} lower than single-sided RowHammer and far double-sided RowHammer, respectively.

We hypothesize that this observation is caused by increased delay to issue \act{} after \pre{} to an aggressor row (i.e., $t_{AggOFF}$~\cite{luo2023rowpress}). 
For example, in the single-sided \comra{} and far double-sided RowHammer access pattern\omcr{1}{s} (\act{}~\src{}-\pre{}-\act{}~\dst{}), the frequency of hammering with \src{} is \ous{relatively lower} than single-sided RowHammer (\act{}~\src{}-\pre{}) due to activating \dst{} in\omcr{1}{-}between every activation of \src{}.
Prior work~\cite{luo2023rowpress} shows that increasing $t_{AggOFF}$ single-sided RowHammer reduces the \gls{hcfirst}.
As a result, single-sided \comra{} and far double-sided RowHammer \one{} exhibit similar \gls{hcfirst} distribution\omcr{1}{s}, and \two{} are more effective than single-sided RowHammer.

\begin{figure}[ht]
    \centering
    \includegraphics[width=\linewidth]{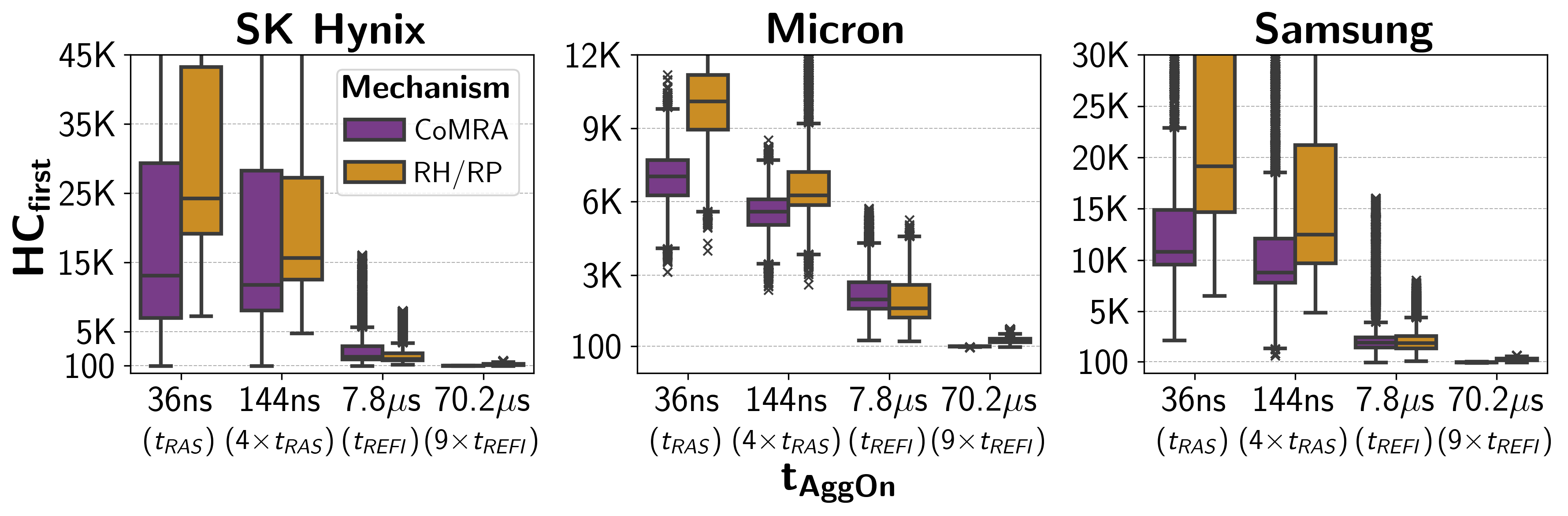}
    \caption{\gls{hcfirst} distribution of double-sided \comra{} and RowPress with different \gls{taggon} values.}
    \label{fig:rc_rp_ds}
\end{figure}

\noindent\textbf{\comra{} vs. RowPress.}
To understand the read disturbance effect of \comra{} better, we analyze how increasing \gls{taggon} (i.e., increasing the latency of \act{}~\dst{}$\rightarrow$~\pre{}) affects \gls{hcfirst} in double-sided \comra{} and compare double-sided \comra{} against double-sided RowPress.
\figref{fig:rc_rp_ds} shows the \gls{hcfirst} distribution across DRAM rows for four \gls{taggon} values: 36ns \ous{($\tras{}$, the nominal timing parameter)}, 144ns (4$\times\tras{}$), 7.8$\mu$s\gls{trefi}, and 70.2$\mu$s (9$\times$\gls{trefi}).

\observation{\ieycr{1}{When performing d}ouble-sided \comra{}, \ieycr{1}{increasing \gls{taggon}} significantly reduces \gls{hcfirst}.}
For example, in Micron \omcr{1}{chips}, \ieycr{1}{\comra{} with \gls{taggon}=70.2$\mu$s leads to a \param{78.74$\times$} reduction in average \gls{hcfirst} compared to \comra{} with \gls{taggon}=36ns.}
We also observe a similar trend for RowHammer/RowPress: when the \gls{taggon} increased to 70.2$\mu$s from 36ns, average \gls{hcfirst} decreases by \param{31.15$\times$}, \ieycr{1}{similar to} prior works~\cite{luo2023rowpress,yaglikci2024svard,olgun2023hbm}.

\observation{At \gls{taggon}=\gls{trefi}, double-sided RowPress becomes more effective than double-sided \comra{}.}

We observe that double-sided \comra{} \ieycr{1}{leads to a reduction in} \gls{hcfirst} \ous{compared to} double-sided RowHammer/RowPress \ous{accross \gls{taggon} values of} 36ns, 144ns, and 70.2$\mu$s.
For example, \ous{at a \gls{taggon} value of} 144ns for Micron \omcr{1}{chips}, average \gls{hcfirst} of \comra{} is \param{1.27$\times$} lower than RowPress.
However, \ous{at} 7.8$\mu$s, RowPress becomes more effective and has \param{1.17$\times$} lower average \gls{hcfirst} than \comra{}.

\takeaway{\ieycr{1}{\emph{Pressing}} with \comra{} \ieycr{1}{is more effective than hammering with \comra{}}.}

\begin{figure}[ht]
    \centering
    \includegraphics[width=\linewidth]{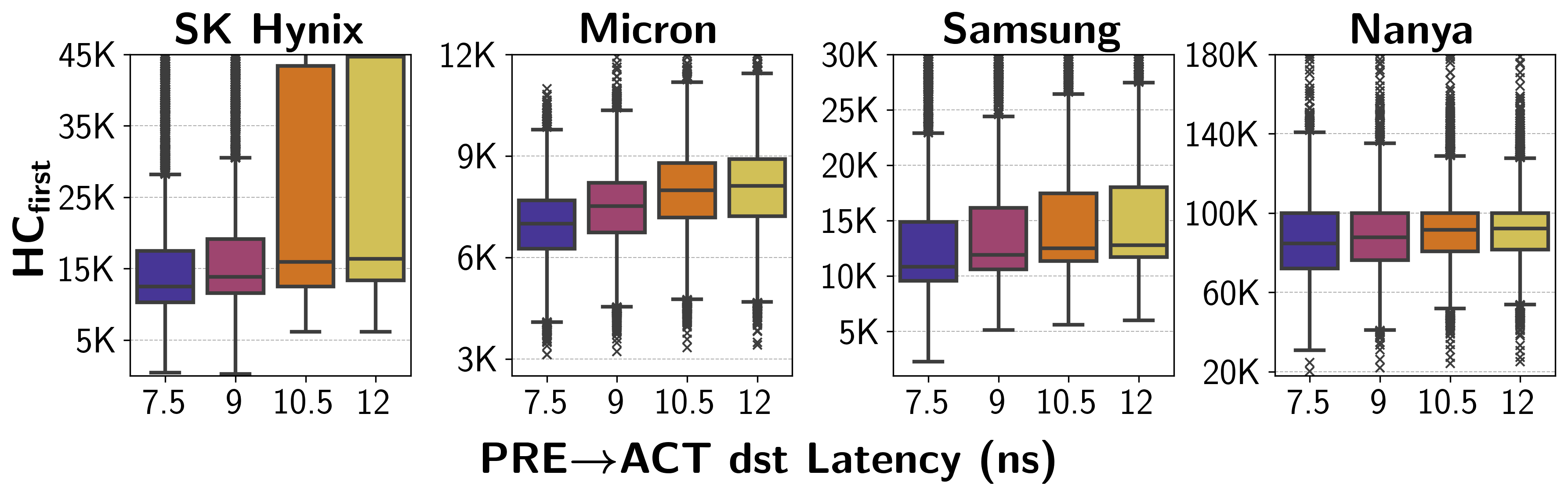}
    \caption{\gls{hcfirst} distribution of double-sided \comra{} for varying numbers of latency.}
    \label{fig:rc_timing} 
\end{figure}

\noindent\textbf{Timing Delay.}
We analyze the \ieycr{1}{violated} latency for \pre{}$\rightarrow$\act{}~\dst{} to develop more insights into \omcr{1}{the} read disturbance effect of \comra{}. 
\figref{fig:rc_timing} shows the \gls{hcfirst} distribution of double-sided \comra{} for four \ieycr{1}{violated} \pre{}$\rightarrow$\act{}~\dst{} latency values. 

\observation{\gls{hcfirst} increases as the latency of \pre{}$\rightarrow$\act{}~\dst{} increases.}
For example, \ous{as the \pre{}$\rightarrow$\act{}~\dst{} latency increases from} 7.5ns to 12ns, average \gls{hcfirst} increases by \param{3.10$\times$}, \param{1.18$\times$}, \param{1.17$\times$}, and \param{3.01$\times$} for SK Hynix, Micron, Samsung, and Nanya.
We hypothesize that \ieycr{1}{the \comra{} access pattern} becomes more of a RowHammer access pattern as the latency increases. As a result, \gls{hcfirst} distribution exhibits higher values when the latency increases.

\begin{figure}[ht]
    \centering
    \includegraphics[width=\linewidth]{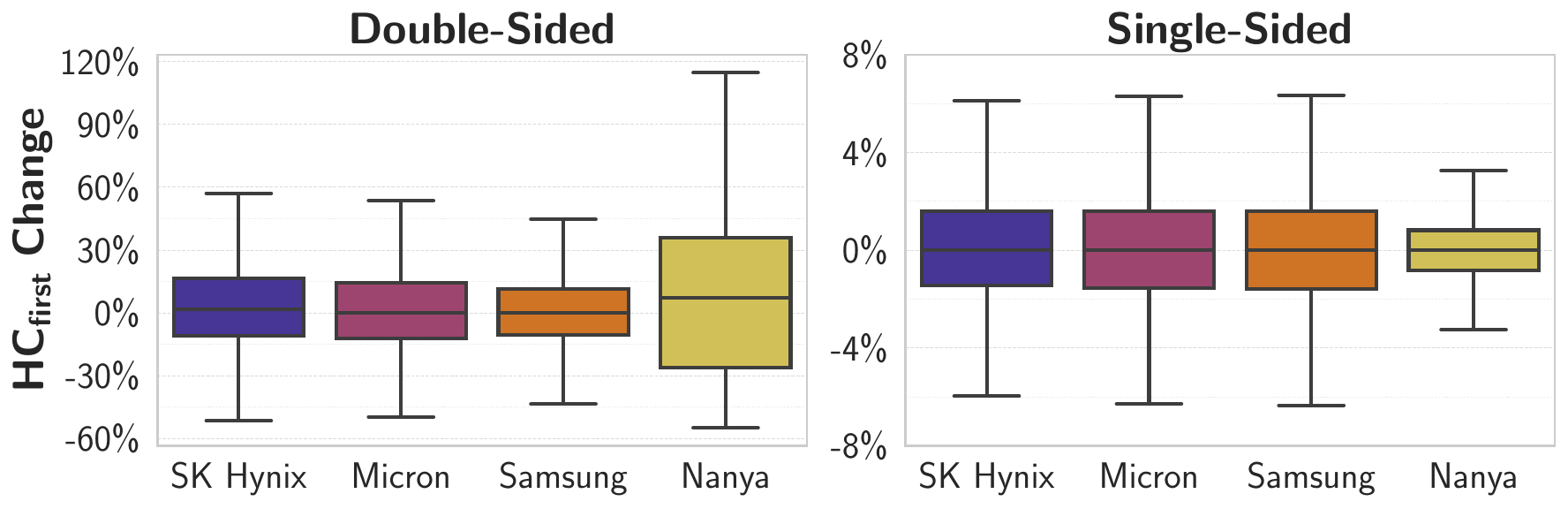}
    \caption{Distribution of the \omcr{1}{change in} \gls{hcfirst} when the copy direction is \omcr{1}{reversed}.}
    \label{fig:rc_access_pattern}
\end{figure}

\noindent\textbf{Copy Direction.}
We analyze how copy direction affects the \gls{hcfirst}. Instead of copying from \src{} to \dst{}, we copy from \dst{} to \src{} and test the \gls{hcfirst} change distribution. \figref{fig:rc_access_pattern} shows the change in \gls{hcfirst} across rows.

\observation{\ous{In most cases}, copy direction has a small effect on \gls{hcfirst}.}
We observe that, across all four manufacturers, the average \gls{hcfirst} change is \param{2.79\%} (i.e., 1.03$\times$) and \param{0.40\%} (i.e., 1.004$\times$) for double-sided and single-sided, respectively. However, in a small fraction of DRAM rows, we observe up to \param{20.10$\times$} and \param{2.39$\times$} change in \gls{hcfirst} for double-sided and single-sided, respectively.

\begin{figure}[ht]
    \centering
    \includegraphics[width=\linewidth]{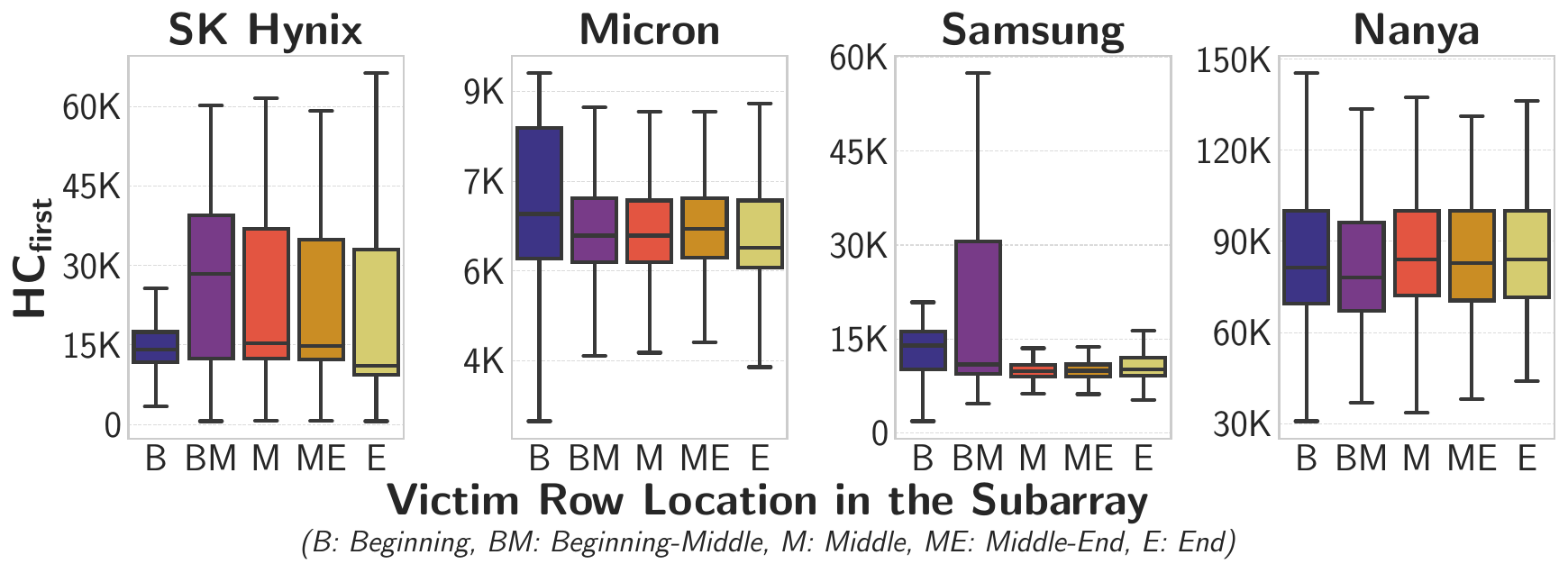}
    \caption{\gls{hcfirst} of double-sided \comra{} based on the victim row's location in a subarray.}
    \label{fig:rc_spatial_var}
\end{figure}

\noindent\textbf{Spatial Variation.}
\figref{fig:rc_spatial_var} shows the \gls{hcfirst} distribution of double-sided \comra{} across DRAM rows (y-axis) based on a victim row's location in a subarray (x-axis).

\observation{\gls{hcfirst} varies based on the victim row's location in a subarray.}
We observe that the physical location of the victim row in the subarray can lead to variations in average \gls{hcfirst} of up to \param{1.40$\times$} for SK Hynix, \param{2.25$\times$} for Micron \ieycr{1}{chips}, \param{2.57$\times$} for Samsung \ieycr{1}{chips}, and \param{1.04$\times$} for Nanya chips.

\observation{Each manufacturer has a different \gls{hcfirst} variation trend \ieycr{1}{with the victim row location in the subarray}.}
For example, in SK Hynix \ieycr{1}{chips}, victim rows at the beginning of a subarray (i.e., the first 20\% rows) exhibit lower average \gls{hcfirst} than others, while in Samsung \ieycr{1}{chips}, victim rows in the middle have the lowest average \gls{hcfirst}. We hypothesize that differences in DRAM circuit design and manufacturing process technology could lead to a different \gls{hcfirst} variation trend for each manufacturer.

\takeaway{\comra{} \one{} decreases \gls{hcfirst} significantly compared to RowHammer, and \two{} gets affected by data pattern, temperature, timing delays, copy direction, and spatial variation.}

\section{Read Disturbance Effect of Simultaneous Multiple-Row Activation (SiMRA) in\\COTS DRAM Chips}
\label{sec:simra}
We present an experimental analysis of \simra{}'s read disturbance effect in \gls{cots} DRAM chips. \simra{} can be used to perform many in-DRAM operations, including
\one{} bulk bitwise operations (e.g., AND, OR,\omcr{0}{NAND, and NOR})~\cite{gao2019computedram, gao2022frac, yuksel2024functionally,yuksel2024simultaneous,olgun2023dram},
\two{} copying data to multiple rows~\cite{yuksel2024simultaneous}, and
\three{} generating true random numbers~\cite{olgun2021quac}.
To understand the read disturbance effect of \simra{}, we repeatedly perform \simra{} to hammer multiple rows simultaneously.

\subsection{Hammering with \simra{}}
\label{subsec:simra_key_idea}

\begin{figure}[ht]
    \centering
    \includegraphics[width=\linewidth]{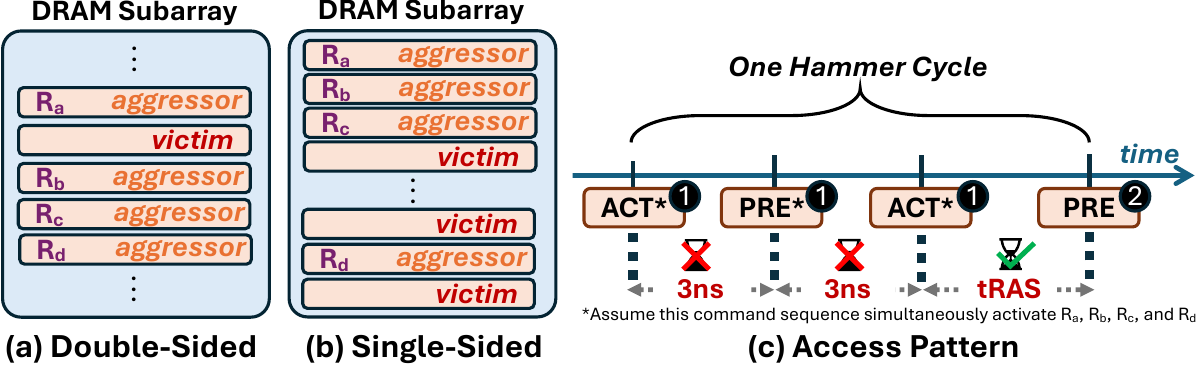}
    \caption{Example of \omcr{0}{(a)} double-sided \simra{} attack, \omcr{0}{(b)} single-sided \simra{} attack, and \omcr{0}{(c)} their access pattern.}
    \label{fig:simra_mech}
\end{figure}

\noindent\textbf{Key Idea.}
We exploit \simra{} to \ous{\emph{simultaneously}} activate multiple rows \ous{and repeatedly} hammer many rows.
To do so, we issue \act{}-\pre{}-\act{} command sequence in quick succession, similar to prior works~\cite{gao2019computedram,olgun2021quac,gao2022frac,olgun2023dram,yuksel2024functionally,yuksel2024simultaneous}.
This results in simultaneously activated rows to be aggressor rows and their neighboring rows to be victim rows.
\figref{fig:simra_mech} illustrates our key idea to perform hammering using \simra{}.
Depending on the location of simultaneously activated rows, we can perform either \one{} a double-sided attack where any two activated rows sandwich a victim row ($R_{a}$ and $R_{b}$ in \figref{fig:simra_mech}a) and \two{} a single-sided attack where activated rows do \emph{not} sandwich a victim row (\figref{fig:simra_mech}b).

\noindent\textbf{Access Pattern \& Operation.}
\figref{fig:simra_mech}c illustrates our access pattern to hammer using \simra{}. Our attack consists of \param{two} key steps. First, we issue a\omcr{0}{n} \act{}-\pre{}-\act{} command sequence with reduced \ous{timing parameters} to simultaneously activate \ous{multiple} aggressor rows, which are $R_{a}$, $R_{b}$, $R_{c}$, and $R_{d}$ in \ous{our} example (\dingOne{}). Second, we wait for $\tras{}$ and issue \omcr{0}{a} \pre{} command to complete the \simra{} operation (\dingTwo{}). We \ous{count} one \simra{} operation as a single hammer \ous{and repeatedly} perform \simra{} \ous{operations} to induce bitflips in victim rows.

\subsection{Experimental Methodology}
\label{subsec:simra_method}
We use the same metrics, algorithms, data patterns, and temperature levels as \comra{} experiments (\secref{subsec:char_comra}). This section explains the rest of the methodology unique to \simra{} experiments.

\noindent\textbf{Finding Simultaneously Activated Rows.} 
Prior works~\cite{olgun2021quac, yuksel2024simultaneous, yuksel2024functionally} show that issuing an \act{}-\pre{}-\act{} command sequence and following with a \wri{} command overwrites the simultaneously activated rows with data \omcr{0}{supplied with the} \wri{} \omcr{0}{command}. We follow the same methodology and reverse engineer the simultaneously activated rows with the \act{}-\pre{}-\act{} command sequence for every row address in a tested subarray. Similar to prior works~\cite{yuksel2024simultaneous, yuksel2024functionally}, we observe that \gls{cots} DRAM chips can activate 2, 4, 8, 16, and 32 rows in the same subarray. \ieycr{0}{We define a term called \emph{\simra{}-N} where N is the number of simultaneously activated rows (i.e., 2, 4, 8, 16, or 32). For example, \simra{}-16 stands for simultaneous activation of 16 rows.}

\noindent\textbf{Timing Delay.}
We sweep two key timing delays in the \act{}-\pre{}-\act{} command sequence in \ous{\figref{fig:simra_mech}c}:~\one{} \omcr{0}{from} \act{} \ous{to} \pre{} and \two{} \omcr{0}{from} \pre{} \ous{to} \act{}.
All experiments are conducted \ous{with timing delays of $3ns$ (as shown in \figref{fig:simra_mech}c)} unless stated otherwise.

\noindent\textbf{Number of Instances Tested.} To maintain a reasonable testing time, we select a total of six subarrays in one bank per DRAM module: two subarrays from the beginning of the bank, two subarrays from the middle of the bank, and two subarrays from the end of the bank. Within each subarray, we randomly test 100 different groups of rows that are simultaneously activated each for 2-, 4-, 8-, 16-,
and 32-row activation.

\subsection{COTS DRAM Chip Characterization}
\label{subsec:simra_char}
This section presents our characterization of the read disturbance effect of \simra{} in SK Hynix chips. While we test all four manufacturers, we note that we do \omcr{0}{\emph{not}} observe \simra{} in Samsung, Micron, and Nanya chips, similar to prior works~\cite{yaglikci2022hira, yuksel2024simultaneous, yuksel2024functionally,gao2022frac,olgun2021quac}.\footnote{\ieycr{0}{Prior works~\cite{yaglikci2022hira,yuksel2024functionally,yuksel2024simultaneous,gao2022frac,olgun2021quac} hypothesize that some chips ignore a DRAM command when the command greatly violates nominal timing parameters.}}

\begin{figure}[ht]
    \centering
    \includegraphics[width=\linewidth]{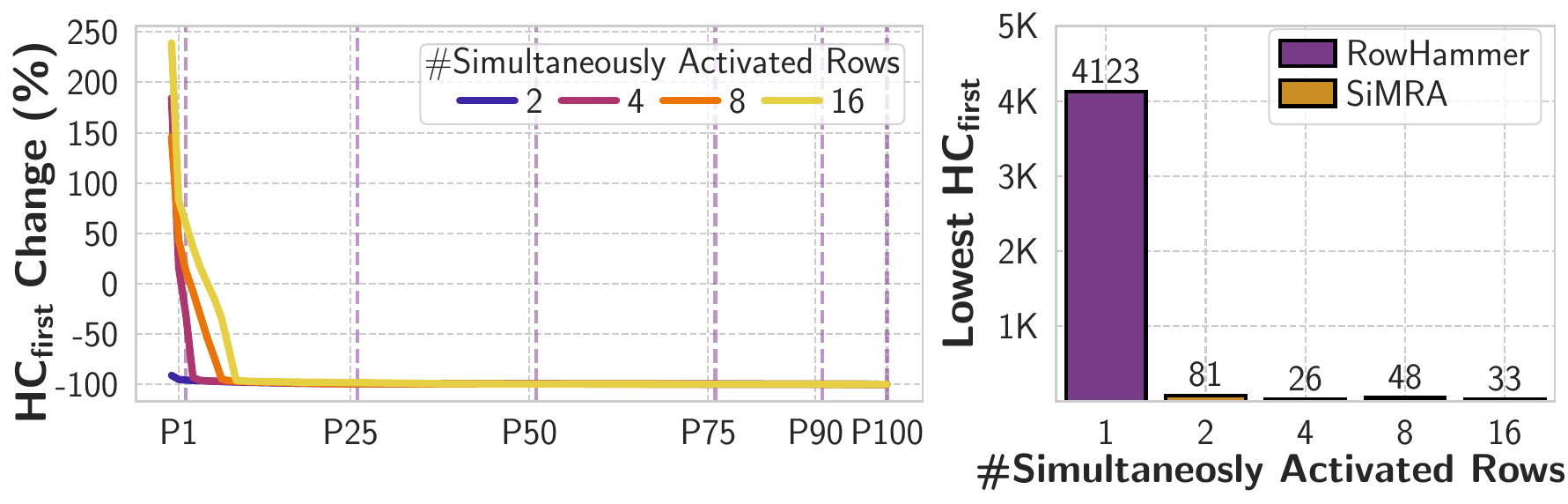}
    \caption{Distribution of the change in \gls{hcfirst} change with double-sided \simra{} compared to double-sided RowHammer (left) and the lowest \gls{hcfirst} observed with double-sided \simra{} and RowHammer (right).}
    \label{fig:simra_rh_ds}
\end{figure}

\noindent\textbf{Double-Sided \simra{} vs. RowHammer.} We investigate the variation in \gls{hcfirst} across DRAM rows when we perform double-sided \simra{} and double-sided RowHammer. In \figref{fig:simra_rh_ds}, the left plot shows the distribution of the change in \gls{hcfirst} (in percentage) when we perform double-sided \simra{} with varying numbers of simultaneously activated rows compared to double-sided RowHammer.\footnote{Even though we simultaneously activate 32 rows, we could not find an activated row group that sandwiches a victim row. Hence, for double-sided \simra{}, we show up to 16-row activation \ieycr{0}{(i.e., \simra{}-16).}} 
The x-axis represents the percentage of all tested victim rows, sorted from the most positive \gls{hcfirst} change to the most negative \gls{hcfirst} change. \figref{fig:simra_rh_ds} (right) shows the lowest \gls{hcfirst} observed across all DRAM chips for double-sided \simra{} with varying numbers of simultaneously activated rows and RowHammer.

\observation{\ous{Hammering with double-sided} \simra{} \omcr{0}{greatly} decreases \gls{hcfirst}.}
\ous{{F}or double-sided \simra{} with 2-, 4-, 8- and 16-row activation, respectively} \param{100.00\%}, \param{98.79\%}, \param{97.40\%}, and \param{94.94\%}, of victim rows \ous{experience lower \gls{hcfirst}} \omcr{0}{than RowHammer}.
At least \param{25.19\%} of victim rows exhibit more than 99\% reduction in \gls{hcfirst}.
We observe that performing double-sided \simra{} \ous{decreases the} \gls{hcfirst} down to \param{26}. 

We observe that the reduction in \gls{hcfirst} is \emph{not} proportional to the number of simultaneously activated rows since at least 25.19\% of victim rows exhibit >99\% reduction (i.e., >100$
\times$reduction) in $HC_{first}$ for \emph{all} tested \ieycr{0}{\emph{N} (i.e., numbers of simultaneously activated rows)}. For example, one tested victim row shows a 158.58x (124.94x) reduction in \hcfirst{} when performing double-sided SiMRA with 4-row activation (32-row activation), which is significantly higher than 4x (32x) \omcr{0}{and non-monotonic with N}.

\takeaway{\simra{} drastically exacerbates DRAM\omcr{0}{'s vulnerability to} read disturbance.}

\begin{figure}[ht]
    \centering
    \includegraphics[width=\linewidth]{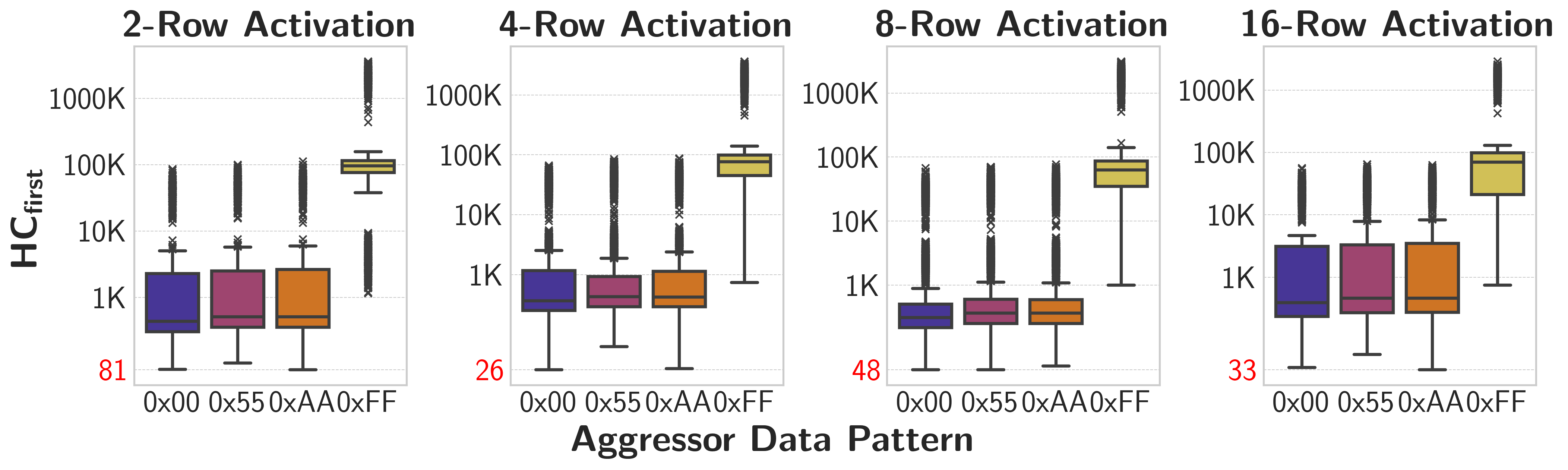}
    \caption{\gls{hcfirst} distribution of double-sided \simra{} for \omcr{0}{different} aggressor data patterns and numbers of activated rows\omcr{0}{.} Victim rows ha\omcr{0}{ve} negated aggressor data pattern.}
    \label{fig:simra_dp}
\end{figure}

\noindent\textbf{Data Pattern.}
We analyze \ous{the effect of data pattern on} \gls{hcfirst} across DRAM rows when we perform double-sided \simra{}.
\figref{fig:simra_dp} shows the \gls{hcfirst} distribution of double-sided \simra{} for four tested data patterns.
Each boxplot is dedicated to a different number of simultaneously activated rows \ieycr{0}{(i.e., N)}.

\observation{Data pattern significantly affects \gls{hcfirst}.}
We observe that, \ous{across all tested \omcr{0}{N values}, initializing the victim rows with a 0x00 data pattern increases} average \gls{hcfirst} by up to \param{57.80$\times$} \ous{when compared to other data patterns}.

\observation{\simra{} and RowHammer have opposite bitflip directions.}
The dominant bitflip direction for \simra{} is 1 to 0 in all tested \omcr{0}{N values} \ieycr{0}{as also shown in Observation 13 that 0xFF as the victim data pattern (i.e., 0x00 as aggressor data pattern) results in much lower \gls{hcfirst} than 0x00 (i.e., 0xFF as aggressor data pattern)}.
For RowHammer (not shown in the figure), we observe that the dominant bitflip direction is 0 to 1, similar to prior work's findings~\cite{luo2023rowpress,luo2024experimental,luo2025revisiting}.

\takeaway{Double-sided \simra{} is significantly affected by data pattern\omcr{0}{. Th}e directionality of \simra{} and RowHammer bitflips \ieycr{0}{are} opposite.}

\begin{figure}[ht]
    \centering
    \includegraphics[width=\linewidth]{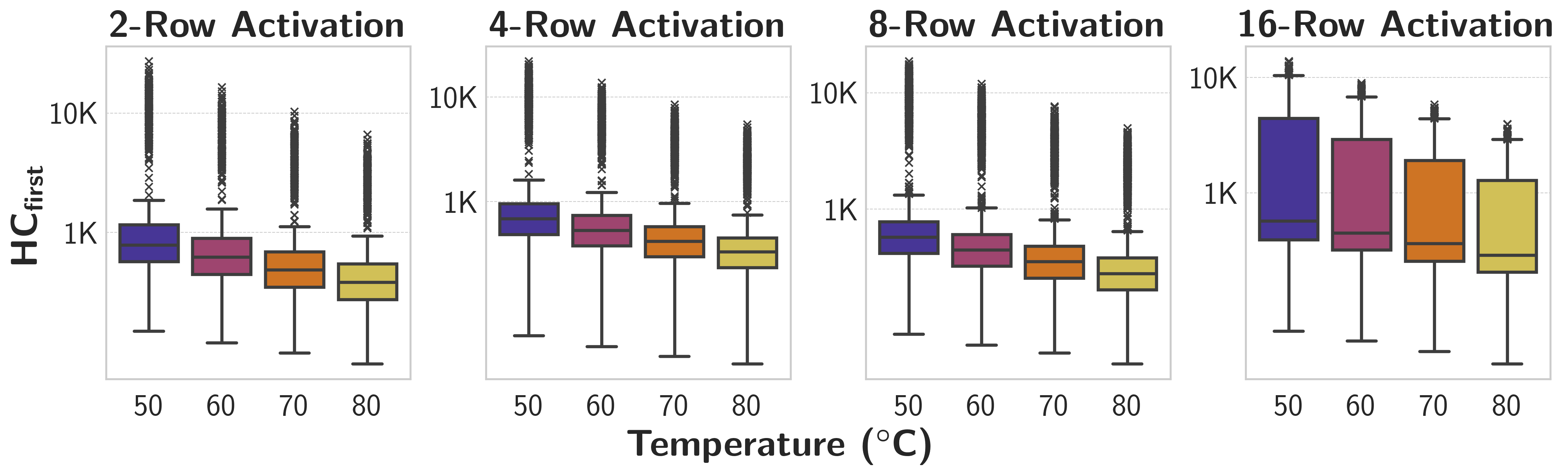}
    \caption{\gls{hcfirst} distribution of double-sided \simra{} at different temperatures \omcr{0}{and} numbers of simultaneously activated rows.}
    \label{fig:simra_temp}
\end{figure}

\noindent\textbf{Temperature.}
\figref{fig:simra_temp} shows the \gls{hcfirst} of double-sided \simra{}\omcr{0}{-N} at four temperature levels: 50$^{\circ}$C, 60$^{\circ}$C, 70$^{\circ}$C, and 80$^{\circ}$C. 

\observation{\gls{hcfirst} decreases as temperature increases.}
Increasing temperature consistently decreases \gls{hcfirst} \omcr{0}{for} all \omcr{0}{N}. For example, from 50$^{\circ}$C to 80$^{\circ}$C, average \gls{hcfirst} decreases by \param{3.24$\times$}, \param{3.10$\times$}, \param{3.02$\times$}, and \param{3.26$\times$} for 2-, 4-, 8-, and 16-row activation.

\ieycr{0}{We hypothesize that double-sided \simra{} has a different underlying silicon-level mechanism compared to double-sided RowHammer due to two observations. First, \simra{} and RowHammer have opposite bitflip directionality. Prior works on both device-level~\cite{zhou2023double, Jie2024Understanding} and real DRAM characterization~\cite{luo2023rowpress, luo2024experimental,luo2025revisiting} show that the dominant bitflip direction for RowHammer is 1 to 0, whereas we observe that for \simra{}, it is 0 to 1 (Observation 14). Second, \simra{} and RowHammer have different temperature dependence. Prior real DRAM characterization works~\cite{luo2023rowpress, orosa2021deeper} show that there is no clear relation between RowHammer and temperature, whereas we observe that for \simra{}, increasing temperature worsens the read disturbance vulnerability (Observation 15).} 
We hope and expect that future device-level studies (inspired by this work) will develop a rigorous device-level understanding of the read disturbance effect of CoMRA and SiMRA operations as device-level studies (e.g.,~\cite{zhou2024Understanding,zhou2024Unveiling}) did for RowPress after the RowPress paper~\cite{luo2023rowpress} demonstrated the empirical basis for the RowPress phenomenon.

\begin{figure}[ht]
    \centering
    \includegraphics[width=\linewidth]{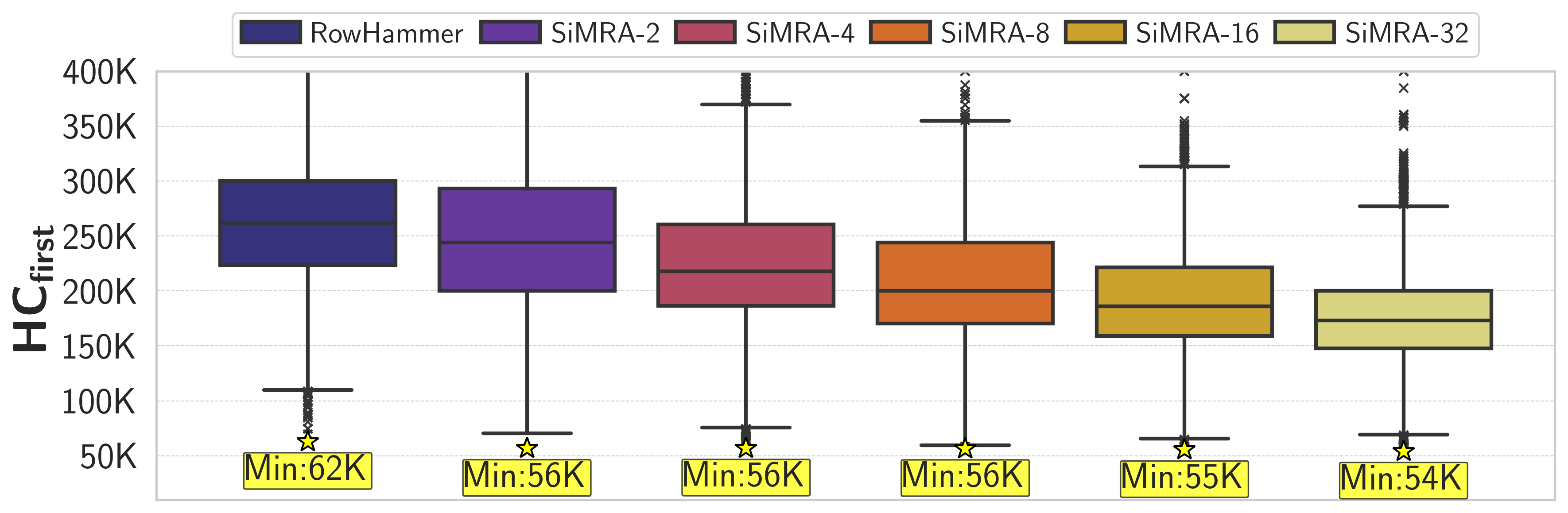}
    \caption{\gls{hcfirst} distribution of single-sided \simra{} with varying numbers of activated rows and RowHammer.}
    \label{fig:simra_ss}
\end{figure}

\noindent\textbf{Single-Sided \simra{}.} \figref{fig:simra_ss} shows the \ieycr{0}{\gls{hcfirst} distribution for} single-sided \simra{}-N and single-sided RowHammer. The lowest \gls{hcfirst} \omcr{0}{of each technique} is highlighted with a yellow rectangle.

\observation{Single-sided \simra{} exhibits \omcr{0}{a} lower \omcr{0}{average \& minimum} \gls{hcfirst} than single-sided RowHammer.}
For example, compared to single-sided RowHammer, the lowest \gls{hcfirst} decreases by \param{1.17$\times$} when performing single-sided \simra{} with 32 rows (i.e., 32-row activation).

\observation{\gls{hcfirst} consistently decreases as the number of simultaneously activated rows increases \ieycr{0}{when performing single-sided \simra{}}.}
For example, average (lowest) \gls{hcfirst} for \simra{}-32 is \param{1.47$\times$} (\param{1.05$\times$}) lower than \simra{}-2.
We hypothesize that \ous{this potentially results} from having more aggressor rows beyond the immediate aggressor of the victim row (i.e., far-aggressor rows) \ous{which} contribute to inducing bitflips in the victim row, \ous{thereby} reducing \gls{hcfirst}, \ous{similarly to the findings of} prior work~\cite{kogler2022half}.

\begin{figure}[ht]
    \centering
    \includegraphics[width=\linewidth]{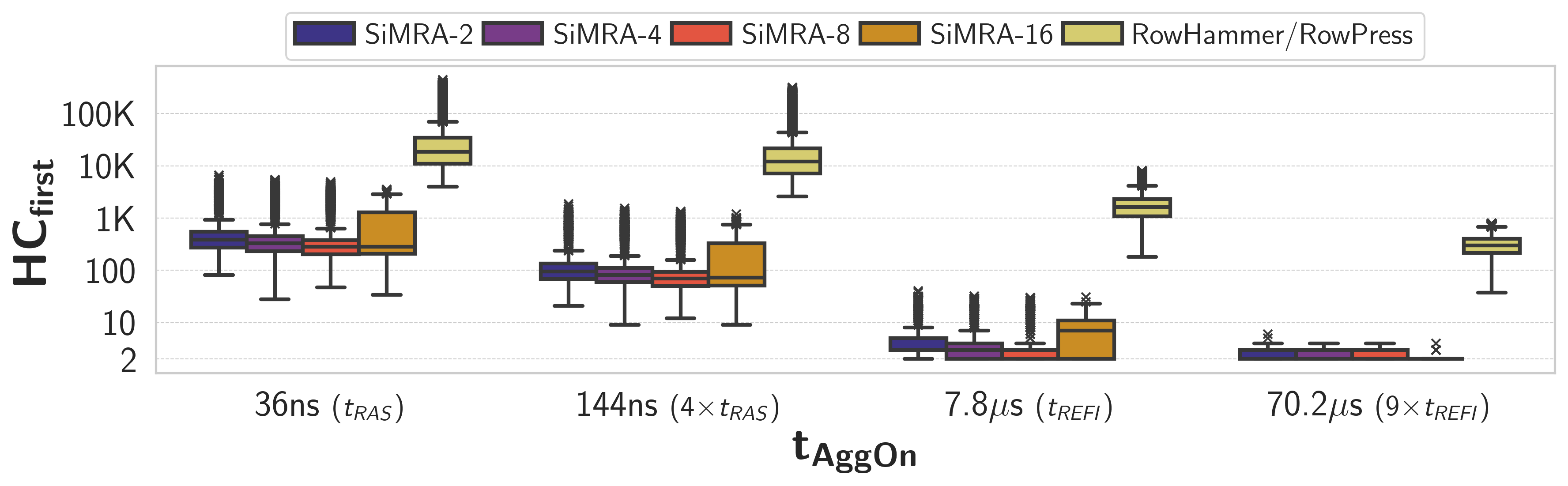}
    \caption{\gls{hcfirst} distribution of RowPress and \simra{} with varying numbers of row activations and \gls{taggon} values.}
    \label{fig:simra_rp_ds}
\end{figure}

\noindent\textbf{\simra{} vs. RowPress.}
To develop a better understanding of \simra{}'s read disturbance effect, we conduct an experiment where we sweep \gls{taggon} for double-sided \simra{}.
We test three \gls{taggon} values \omcr{0}{in addition to} $\tras{}$ \ous{(four in total)} after issuing \act{}-\pre{}-\act{} commands, which keeps the simultaneously activated rows open for longer times.
\figref{fig:simra_rp_ds} shows the \gls{hcfirst} distribution of double-sided \simra{} and RowPress (\ous{we refer to RowPress at 36ns or $\tras{}$ as RowHammer}).

\observation{\gls{hcfirst} decreases \omcr{0}{greatly} as \gls{taggon} increases.}
For example, \ous{when \gls{taggon} increases from 36ns to 70.2$\mu$s \omcr{0}{with} \simra{}}, average \gls{hcfirst} decreases by between \param{144.93$\times$}-\param{270.27$\times$} \ous{across all numbers of simultaneously activated rows}.

\begin{figure}[ht]
    \centering
    \includegraphics[width=\linewidth]{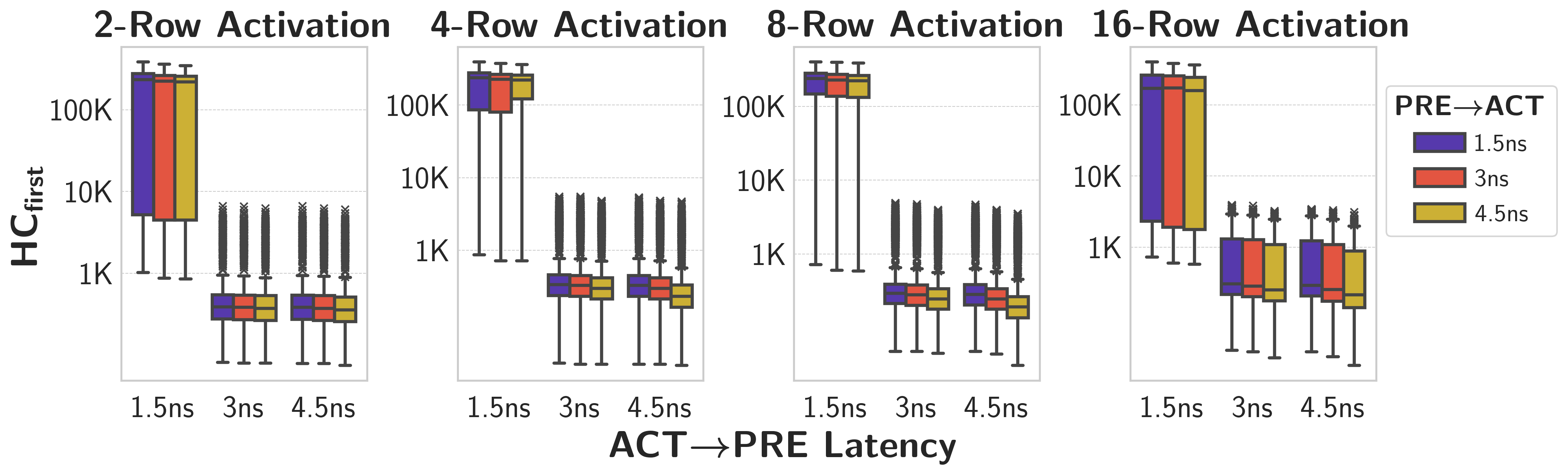}
    \caption{\gls{hcfirst} distribution of double-sided \simra{} for \omcr{0}{different} timing delays \ieycr{0}{between \act{}-\pre{} and \pre{}-\act{}}.}
    \label{fig:simra_timing}
\end{figure}

\noindent\textbf{Timing Delay.} \figref{fig:simra_timing} shows how \gls{hcfirst} changes when we sweep timing delays between \act{}-\pre{} and \pre{}-\act{} in \act{}-\pre{}-\act{}.

\observation{Increasing \pre{}-\act{} latency slightly decreases \gls{hcfirst}.}
\ieycr{0}{For example, \simra{}-16 with \act{}$\rightarrow$\pre{}=3ns, average \gls{hcfirst} decreases by \param{1.23$\times$} when \pre{}$\rightarrow$\act{} increases from 1.5ns to 4.5ns.}

\observation{Specific latency values fail to fully activate aggressor rows, \omcr{0}{which leads to} increase \ieycr{0}{in} \gls{hcfirst}.}
Choosing \act{}$\rightarrow$\pre{} latency as 1.5ns, increases average \gls{hcfirst} by \param{2.28$\times$}. \ieycr{0}{We observe that (similar to prior work~\cite{yuksel2024simultaneous}) some aggressor rows are not fully activated (i.e., not all cells are activated) in very low latencies. We hypothesize that due to this partial activation of aggressor rows leads to a drastic increase in \gls{hcfirst} in some cases.}

\begin{figure}[ht]
    \centering
    \includegraphics[width=\linewidth]{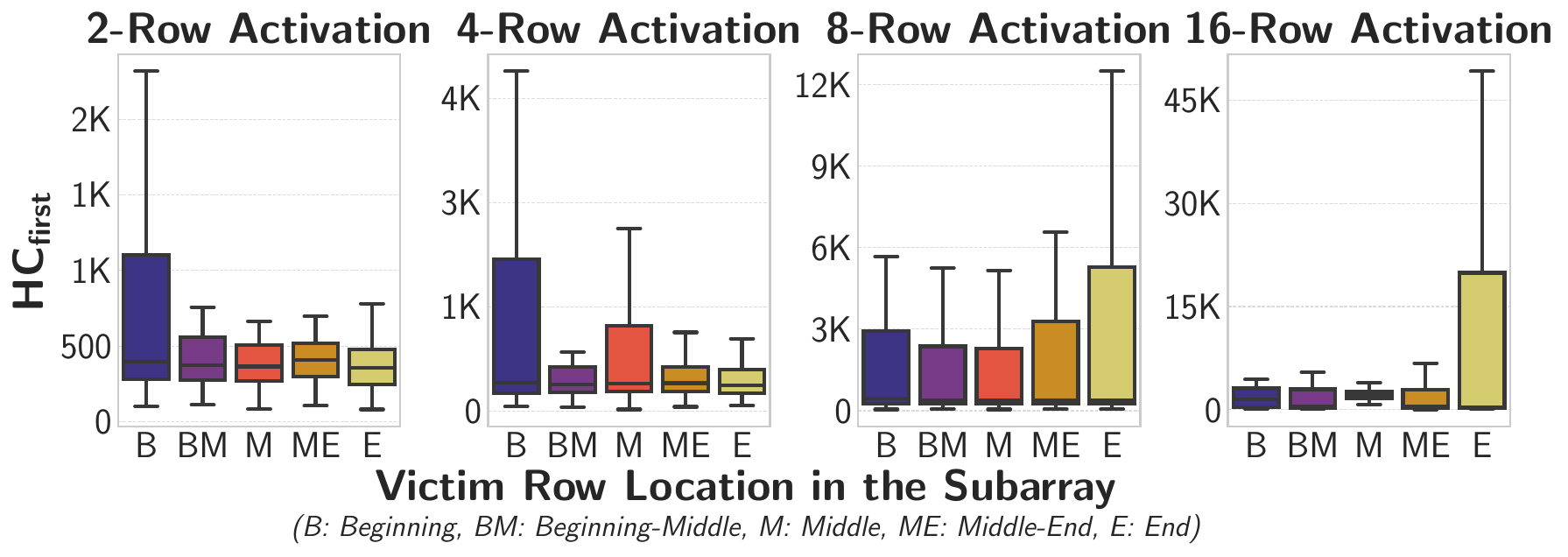}
    \caption{\gls{hcfirst} distribution of double-sided \simra{} based on the victim row's location in a subarray.}
    \label{fig:simra_spatial}
\end{figure}

\noindent\textbf{Spatial Variation.} \figref{fig:simra_spatial} shows the \gls{hcfirst} distribution of double-sided \simra{} (y-axis) with varying numbers of activated rows (each subplot) based on a victim row's location in a subarray (x-axis).

\observation{The effectiveness of \simra{} depends on the victim row's location in a subarray.}
\gls{hcfirst} varies across different victim row locations in a subarray, and this variation is different for each number of simultaneously activated rows.
For example, for 4-row activation, victim rows at the beginning of a subarray experience the highest \gls{hcfirst} distribution, whereas for 8-row activation, victim rows at the end of a subarray exhibit the highest \gls{hcfirst} distribution. 

\takeaway{The operating parameters (i.e., data pattern, temperature, timing delays, and spatial variation) impact \simra{}'s read disturbance effect.}

\section{Combined Read Disturbance Effect of RowHammer with PuD in COTS DRAM Chips}
\label{sec:combined}
We demonstrate how combining RowHammer with \comra{} and \simra{} affects read disturbance vulnerability in real DRAM chips.
To understand how much \ous{the observed \gls{hcfirst} using} RowHammer \ous{decreases when combined with} multiple-row activation:
\one{} we hammer a victim row using multiple-row activation for a fixed \ous{number of times} and
\two{} we perform RowHammer.\footnote{\ieycr{0}{Our tested access pattern is one of many ways of combining the \comra{}, \simra{}, and RowHammer access patterns. There could be more effective access patterns that reduce \gls{hcfirst} even more. We leave this exploration, which requires extensive characterization and analysis, to future work.}}

\subsection{Experimental Methodology}
\label{subsec:combined_method}
\begin{figure}[ht]
    \centering
    \includegraphics[width=\linewidth]{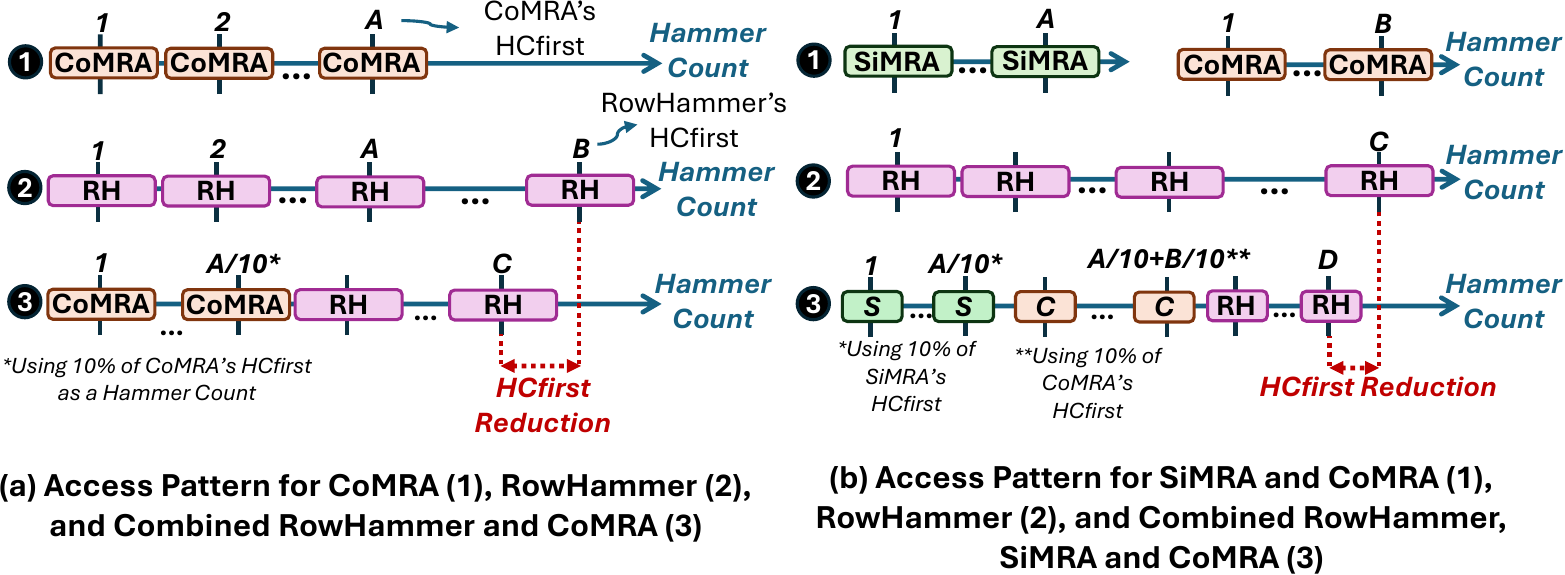}
    \caption{Example access pattern of combining RowHammer (a) with CoMRA and (b) CoMRA and SiMRA together.}
    \label{fig:acc_patt}
\end{figure}

\ous{\figref{fig:acc_patt} shows how RowHammer can be combined with a \omcr{0}{one} (\figref{fig:acc_patt}a presents an example with \comra{}) or two (\figref{fig:acc_patt}b) multiple-row activation techniques.
We combine RowHammer with a single multiple-row activation technique in two steps.
First, we characterize the \gls{hcfirst} of a row when only hammered with a multiple-row activation technique (e.g., \figref{fig:acc_patt}a-\dingOne{} for \comra{}).
\iey{Second, we characterize the \gls{hcfirst} of a victim row with only RowHammer (e.g., \figref{fig:acc_patt}a-\dingTwo{})}
\iey{Third}, we hammer the victim row to a fraction of the multiple-row activation's \gls{hcfirst} value. We use three levels of hammer count for our experiments: 10\%, 50\%, and 90\% of multiple-row activation's \gls{hcfirst} value. For example, if a victim row has \gls{hcfirst} of $A$ when performing \comra{} only (\figref{fig:acc_patt}-\dingOne{}) and if we are to test for 10\%, we hammer that victim row for A/10 times with \comra{}, as in \figref{fig:acc_patt}a.
\iey{Fourth,} we continue hammering the victim row with RowHammer until \iey{the first} bitflip is observed (e.g., \figref{fig:acc_patt}a-\dingThree{}).
We combine RowHammer with both multiple-row activation techniques similarly by characterizing \gls{hcfirst} values (\figref{fig:acc_patt}b-\dingOne{}) and hammering a victim row with each technique \omcr{0}{up} to a fraction and then continuing with RowHammer until a bitflip is observed (\figref{fig:acc_patt}b-\dingThree{})}.
\iey{Fifth,} we report the \gls{hcfirst} change of these combined \omcr{0}{access patterns} compared to using RowHammer only (e.g., B-C decrease in \figref{fig:acc_patt}a and C-D decrease in \figref{fig:acc_patt}b).

\subsection{COTS DRAM Chip Characterization}
\label{subsec:combined_char}
\omcr{0}{We} present our characterization of the combined RowHammer and multiple-row activation pattern in the same chips that are used in \simra{} characterization (\secref{sec:simra}). We conduct our experiments at 80$^{\circ}$C using the double-sided pattern for all techniques with WCDP for each DRAM row. 

\begin{figure}[ht]
    \centering
    \includegraphics[width=\linewidth]{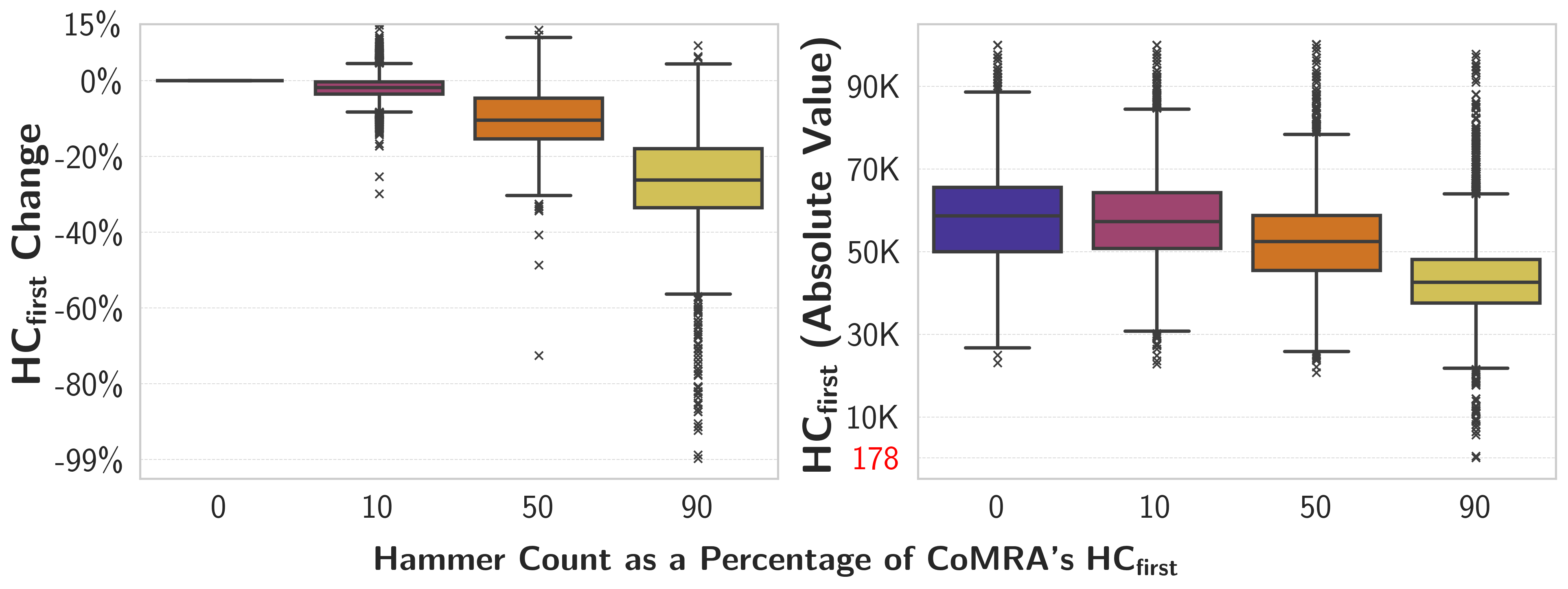}
    \caption{\ieycr{0}{Change in \gls{hcfirst} (left) and absolute \gls{hcfirst} values (right) when combining RowHammer and \comra{}. Hammer count of 0\% represents performing RowHammer only.}}
    \label{fig:combined_rc_rh}
\end{figure}

\noindent\textbf{Combining RowHammer with \comra{}.}
\figref{fig:combined_rc_rh} shows the change in \gls{hcfirst} distribution \ieycr{0}{and the absolute \gls{hcfirst} values} across DRAM rows when \omcr{0}{we} perform combined double-sided RowHammer and \comra{}, compared to double-sided RowHammer.
The x-axis shows the hammer count for \comra{} as a percentage of \gls{hcfirst} observed for each row.

\observation{DRAM rows experience significantly lower \gls{hcfirst} when performing combined RowHammer and \comra{}.}
Compared to double-sided RowHammer, \param{95.33\%} of \ous{tested} victim rows show lower \gls{hcfirst} when we combine double-sided RowHammer with \comra{}.
\gls{hcfirst} reduc\omcr{0}{tion increases} as we increase the \ous{hammering fraction of} \comra{}.
\ous{For example, when a victim row is first hammered with \comra{} until 90\% and 10\% of \comra{}'s \gls{hcfirst} and then hammered with RowHammer, \gls{hcfirst} decreases by \param{1.34$\times$} and \param{1.02$\times$} \omcr{0}{(compared to RowHammer)}, respectively.}

\begin{figure}[ht]
    \centering
    \includegraphics[width=\linewidth]{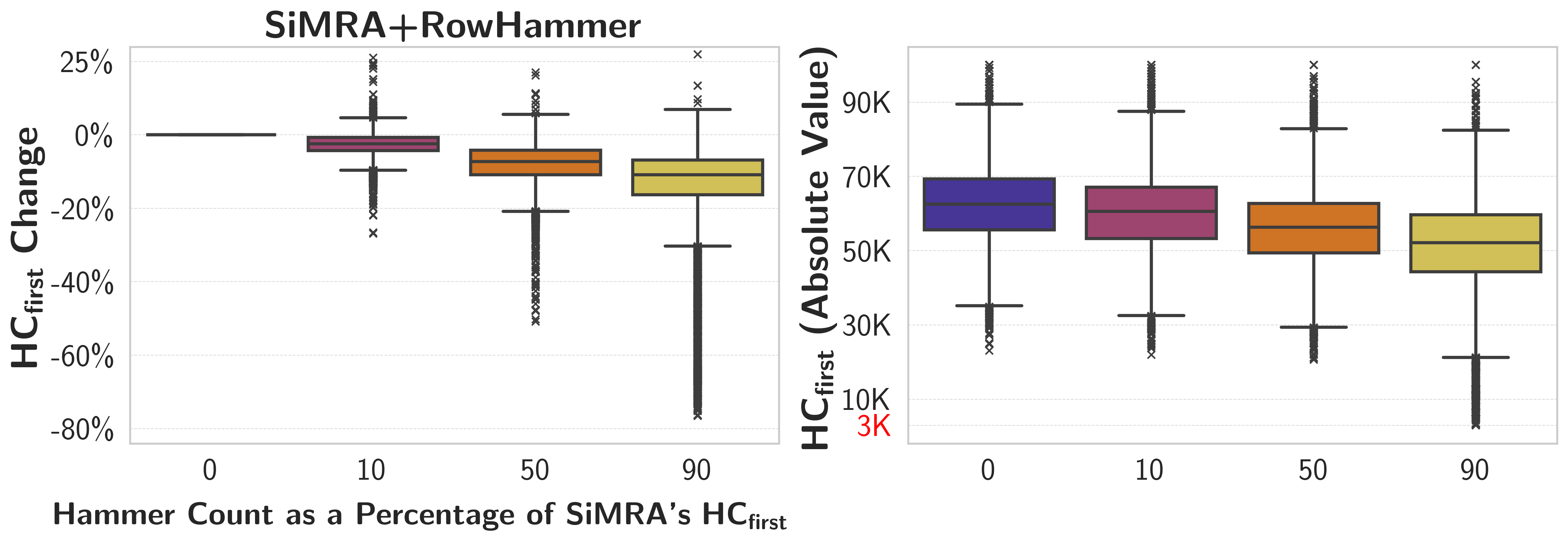}
    \caption{\ieycr{0}{Change in \gls{hcfirst} (left) and absolute \gls{hcfirst} values (right) when combining RowHammer and \simra{}. Hammer count of 0\% represents performing RowHammer only.}}
    \label{fig:combined_simra_rh}
\end{figure}

\noindent\textbf{Combining RowHammer with \simra{}.} \figref{fig:combined_simra_rh} shows the change in \gls{hcfirst} \ieycr{0}{and the absolute \gls{hcfirst} values} when \omcr{0}{we} perform combined double-sided RowHammer and \simra{}. 

\observation{DRAM rows tend to experience lower \gls{hcfirst} when performing combined RowHammer and \simra{}.}
Combining RowHammer with \simra{} \one{} decreases \gls{hcfirst} as the hammer count for \simra{} increases and \two{} is less effective than combining RowHammer with \comra{}. For example, in combined RowHammer and \simra{}, the average \gls{hcfirst} change in the highest hammer count (i.e., 90\% percentage) is \param{1.22$\times$} lower than combined RowHammer and \comra{}. We hypothesize that the most vulnerable cell \ieycr{0}{to RowHammer} in some victim rows is not vulnerable to \simra{}. Thus, combining RowHammer with \simra{} does not decrease the \gls{hcfirst} of all tested victim rows as much as combining \omcr{0}{RowHammer} with \comra{}.

\begin{figure}[ht]
    \centering
    \includegraphics[width=\linewidth]{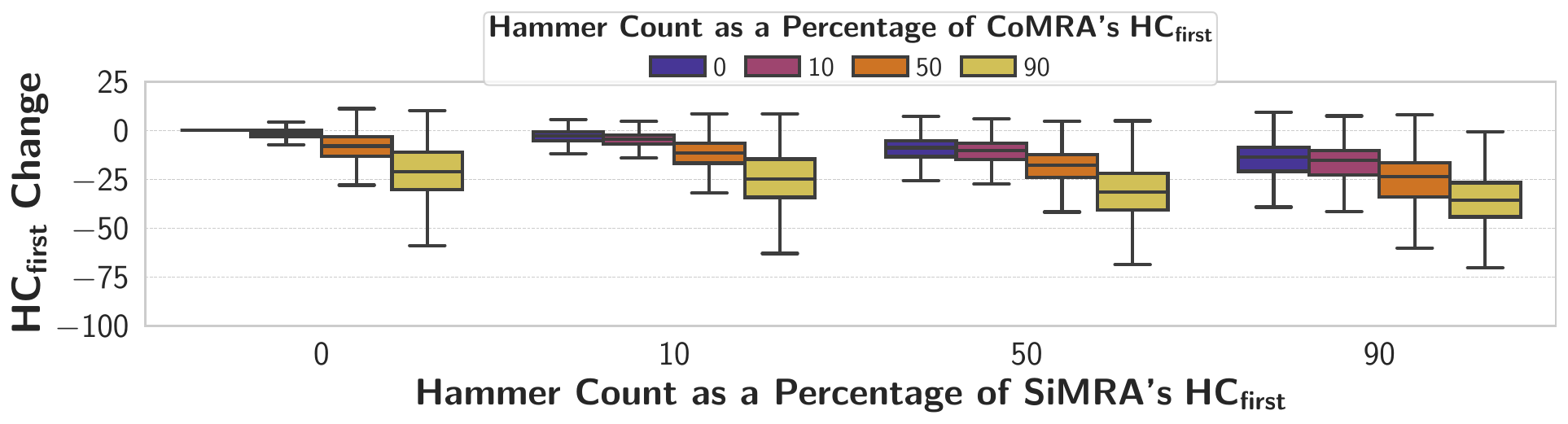}
    \caption{Change in \gls{hcfirst} for combined RowHammer, \comra{}, and \simra{} compared to only RowHammer.}
    \label{fig:mra_rc_rh}
\end{figure}

\noindent\textbf{Combining RowHammer with \comra{} and \simra{}.} \figref{fig:mra_rc_rh} shows the change in \gls{hcfirst} when \omcr{0}{we} combin\omcr{0}{e} double-sided RowHammer with \comra{} and \simra{} together (e.g., \figref{fig:acc_patt}b-\dingThree{}).

\observation{Combined RowHammer with \comra{} and \simra{} together is the most effective combined access pattern to decrease \gls{hcfirst}.}

We observe that \ieycr{0}{when we combine RowHammer with \comra{} and \simra{}}, the majority of the DRAM rows exhibit much lower \gls{hcfirst} \ieycr{0}{compared to performing only RowHammer}. The minimum average \gls{hcfirst} of combined RowHammer with \comra{} and \simra{} together is \param{1.66$\times$} lower than performing only RowHammer.

\takeaway{Combining RowHammer with \comra{} and/or \simra{} decrease\omcr{0}{s} \gls{hcfirst}, and combining RowHammer with \comra{} and \simra{} together is the most effective pattern across all tested \omcr{0}{access} patterns.}

\section{PuDHammer in the Presence\\of \omcr{0}{I}n-DRAM TRR}
\label{sec:trr}
{To prevent read disturbance bitflips, DRAM manufacturers equip their chips with a mitigation mechanism broadly referred to as Target Row Refresh (TRR)~\cite{hassan2021utrr,frigo2020trrespass,micron2016trr,zhang2022softtrr,marazzi2022protrr}. Manufacturers do \emph{not} disclose the operational principles or implementations of proprietary TRR versions (e.g., in DDR4~\cite{hassan2021utrr,frigo2020trrespass}). At a high level, TRR 1) identifies potential aggressor rows as the memory controller issues activate commands to the DRAM chip and 2) preventively refreshes victim rows when the memory controller issues a periodic \texttt{REF} command.}
{We demonstrate that in a tested SK Hynix DDR4 DRAM module~\cite{hynixtrr}, both \comra{} and \simra{} bypass the TRR mechanism and induce more read disturbance bitflips than RowHammer.}

\noindent\textbf{{Uncovering the TRR Mechanism.}}
{We uncover the TRR mechanism in the tested module using U-TRR~\cite{hassan2021utrr,hassan2021utrrgithub}. We observe that the tested module uses a sampling-based TRR mechanism, where TRR probabilistically identifies one aggressor row by sampling row addresses of the last 450 \texttt{ACT} commands before issuing the TRR-capable \texttt{REF} (i.e., \texttt{REF} command that refreshes victim rows).}

\noindent\textbf{{Access Pattern.}}
{We use the custom access pattern reported \omcr{0}{by} U-TRR~\cite{hassan2021utrr} for RowHammer and \comra{}. The {custom} access pattern {(which we call the N-sided pattern)} uses N aggressor rows, (we sweep N from 1 to 10) and one dummy row. Our custom access pattern consists of four key steps. First, we perform a total of 156 hammers for N aggressor rows in one periodic refresh window (see \secref{sec:dram_organization}).\footnote{{The memory controller issues a \texttt{REF} once every 7.8$\mu$s. This allows at most 156 \texttt{ACT} commands to a single DRAM bank in the tested module in a refresh window.}} Second, we hammer the dummy row 468 times (i.e., 156$\times$3 times, the maximum number of \texttt{ACT} commands that can be issued in three refresh windows) to make the TRR mechanism refresh the victims of the dummy row while activations to the N aggressor rows go unnoticed.}
{For \simra{}, instead of using N aggressor rows, we perform \simra{} (described in \secref{sec:simra}) that simultaneously activates multiple (2, 4, 8, 16, and 32) rows. We hammer each aggressor row 500K times. We repeat this test 5 times and report the average, maximum, and minimum bitflip \omcr{0}{counts} across all tests.} 

\begin{figure}[ht]
    \centering
    \includegraphics[width=0.97\linewidth]{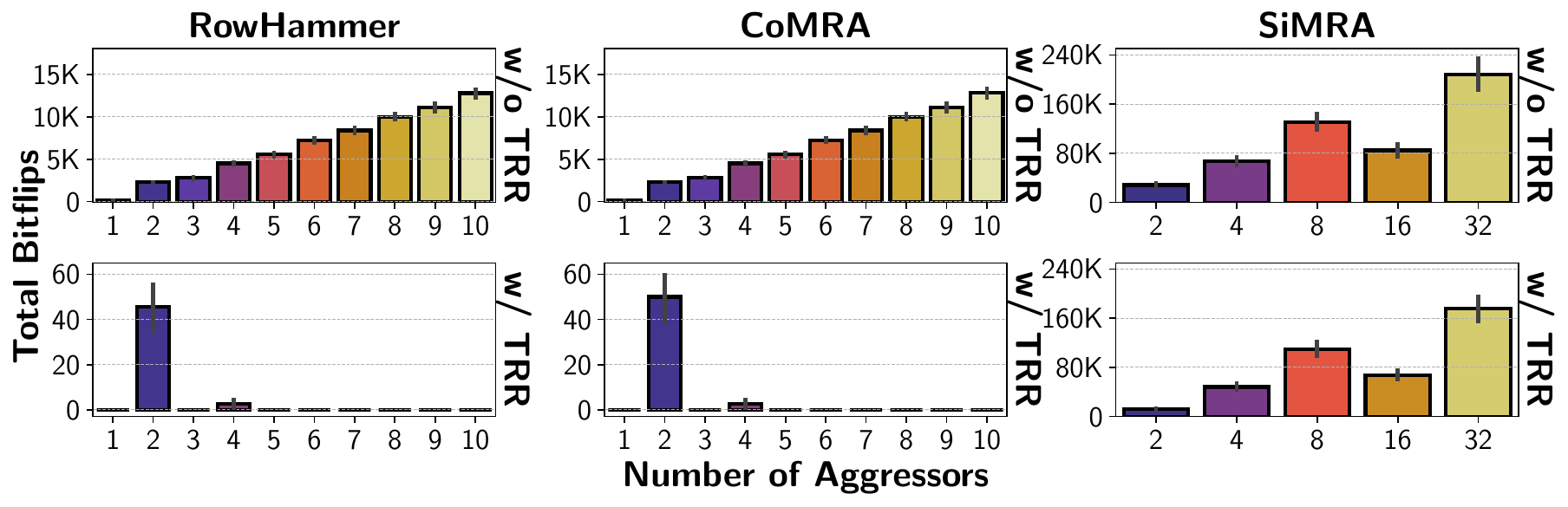}
    \caption{{Number of bitflips in victim rows averaged across all tests when performing RowHammer, \comra{}, and \simra{} without TRR (top) and with TRR (bottom).}}
    \label{fig:many_sided}
\end{figure}

\noindent\textbf{{Results.}} 
{\figref{fig:many_sided} shows the number of bitflips observed in victim rows on average across all tests, with error bars representing the range across all tests. The top plot shows the results when TRR is disabled (i.e., w/o TRR), and the bottom subplot shows the results when TRR is enabled (i.e., w/ TRR). Each column of subplots is dedicated to a different hammering technique.}

\observation{{\simra{} and \comra{} bypass the TRR mechanism and induce more bitflips than RowHammer.}}

{For example, \simra{} with 32-row activation and 2-sided \comra{} respectively induce 11340$\times$ and 1.10$\times$ more bitflips than 2-sided RowHammer on average \ieycr{0}{across all 5 iterations}.}

\observation{{TRR cannot effectively mitigate \simra{}-induced read disturbance bitflips.}}

{{We observe that \simra{} bitflips reduce only \param{15.62}\% on average \omcr{0}{with TRR}, whereas RowHammer and \comra{} bitflips reduc\omcr{0}{e} greatly \ieycr{0}{(e.g., \param{99.89}\% reduction on average for RowHammer}).} We hypothesize that this is due to two reasons. First, since \simra{} only issues two \act{} commands back-to-back (i.e., ACT-PRE-ACT, see \secref{subsec:back_pud} and \secref{subsec:simra_key_idea}) to simultaneously activate up to 32 rows, the TRR mechanism likely can only mitigate the bitflips in the victims of the two aggressor row addresses issued with the two ACT commands. Second, since the \gls{hcfirst} of \simra{} is much lower (e.g., 26) than the maximum number of ACT commands we can issue in a periodic refresh window (i.e., 156), \simra{} induces read disturbance bitflips before \omcr{0}{the memory controller can} issue a TRR-capable REF command.}

\takeaway{{\simra{} and \comra{} bypass in-DRAM TRR mechanism and are more effective than RowHammer to induce read disturbance bitflips.}}

\section{{Mitigating PuDHammer}}
\atb{We experimentally demonstrate the potential interaction between 
read disturbance and multiple-row activation-based \gls{pud} mechanisms.
We perform multiple-row activation operations within the constraints of current DRAM chips.
Today's DRAM chips are \emph{not} designed to support these operations.}
Our \agy{results provide insights into read 
disturbance challenges that \gls{pud}-enabled systems might face and show that read disturbance vulnerability 
should be taken into account while designing \gls{pud}-enabled systems}.\footnote{Future solutions to such challenges should include designing DRAM chips to fundamentally 1)~support CoMRA and SiMRA and 2)~mitigate read disturbance. Our analysis aids the development of such designs by providing new experimental insights.}

{Future work should investigate countermeasures in detail (based on specific PuD designs) and propose further countermeasures based on our findings. We leave a detailed investigation of such analyses and countermeasures for future work, but 1) sketch three types of}
\agy{countermeasures against PuD-based read disturbance bitflips {and 2) adapt and evaluate the state-of-the-art RowHammer mitigation standardized by industry, called Per Row Activation Counting (PRAC)~\cite{saroiu2024ddr5, jedec2024jesd795c, canpolat2025chronus,canpolat2024understanding}, to account for PuDHammer.}}

\subsection{{Countermeasures Against PuDHammer}}
\label{sec:countermeasures}
\head{\agy{Separating PuD-enabled rows.}}
\agy{Prior PuD architectur\atb{es}~\cite{seshadri2017ambit, hajinazar2021simdram, deoliveira2024mimdram,ali2019memory} split a DRAM array into two parts, where a small set of rows are used for computation \ous{(compute region)} and the remaining large fraction of rows for storing data \ous{(storage region)}.
\atb{In such an architecture, \comra{} and \simra{} operations would happen 
repeatedly in the compute region, resulting in activation counts exceeding
what we experimentally demonstrated to cause read disturbance bitflips
(e.g., to perform \omcr{0}{an} 8-bit multiplication operation, SIMDRAM~\cite{hajinazar2021simdram} issues 663 \comra{}
and \simra{} operations).}
Future PuD systems can be designed with two \hluo{constraints}:
1)~\simra{} operations are allowed \emph{only} on the compute \ous{region} rows, and
2)~\iey{at} most one of the operands in \comra{} operations (source and destination rows) can be a row outside of the compute region.

These two \hluo{constraints}
1)~\ous{limit} \simra{}'s and double-sided \comra{}'s negative impact on read disturbance to the compute region and
2)~\ous{limit the \gls{hcfirst} decrease on storage region rows} to the \ous{reduction} of single-sided \comra{} operations.
To prevent read disturbance bitflips in the compute region, each row in the relatively small compute region (e.g., \param{3 to 32} rows~\cite{hajinazar2021simdram, seshadri2017ambit, deoliveira2024mimdram,yuksel2024simultaneous,ali2019memory} out of 1024 rows in a DRAM subarray~\cite{yuksel2024functionally, yuksel2024simultaneous, yaglikci2022hira, olgun2024read, olgun2023hbm, patel2017reach}) can be \ous{periodically} refreshed after a number of \simra{} operations (e.g., 20).
These refresh operations can be spread over time similar\omcr{0}{ly} to how periodic refreshes are performed (\iey{i.e.,} after each \simra{} operation, a portion of rows in the compute region are refreshed).
To prevent read disturbance bitflips outside the compute region, existing RowHammer mitigation mechanisms can be simply configured for a reduced \gls{hcfirst} value as the reduction due to single-sided \comra{} is less than \param{2\%} \ieycr{0}{(\figref{fig:rc_rh_ss})}.
Doing so is possible at low hardware complexity as \ieycr{0}{this approach} does \emph{not} require sophisticated tracking mechanisms, \iey{however,} it might cause performance and energy overheads.}

\head{\agy{Weighted contribution of different row activation types.}}
\agy{Based on the \gls{hcfirst} reduction factors of \comra{} and \simra{}, each \comra{} or \simra{} operation can be accounted for an equivalent hammer count using double-sided hammering, e.g., assuming that \comra{} reduces \gls{hcfirst} by 20$\times$, each \comra{} operation can increment row activation counters by 20.
Prior works~\cite{qureshi2024impress, luo2023rowpress} propose a similar approach to prevent RowPress bitflips by slightly modifying existing RowHammer mitigation mechanisms.}

\head{\agy{\omcr{0}{Clustered} multi\iey{ple-}row activation.}} 
\ous{We demonstrate that double-sided \simra{} significantly reduces \gls{hcfirst} (\secref{sec:simra}).
This operation is possible because the row decoder circuitry simultaneously activates \iey{multiple} rows in a variety of locations within the subarray (as shown in prior work~\cite{yuksel2024simultaneous}), thereby causing some of the simultaneously activated rows to sandwich an unactivated row.
Future PuD-enabled systems can mitigate double-sided \simra{} by employing row decoder circuits that \omcr{0}{cluster} simultaneous row activation by guaranteeing \omcr{0}{that adjacent rows are activated, i.e., there are} no sandwiched unactivated rows.}

\subsection{{Adapting Existing RowHammer Mitigations}}
\label{subsec:rh_mitigation}
{In this section, we adapt and evaluate the state-of-the-art industry solution to RowHammer (i.e., PRAC~\cite{saroiu2024ddr5, jedec2024jesd795c, canpolat2024understanding, canpolat2025chronus}) to mitigate PuDHammer. We demonstrate that \omcr{0}{adapted} PRAC incurs 48.26\% average system performance overhead across all tested workloads.}

\noindent\textbf{{PRAC Overview.}}
{
PRAC~\cite{saroiu2024ddr5, jedec2024jesd795c, canpolat2025chronus,canpolat2024understanding}, as described in the JEDEC DDR5 standard updated in April 2024~\cite{jedec2024jesd795c}, implements a counter for each row. 
Upon an \texttt{ACT}, PRAC increases the counter of the activated row and thus accurately measures the activation counts of all rows.
When a row’s activation count reaches a threshold (the read disturbance threshold (RDT) is the minimum number of activations needed to induce a read disturbance bitflip in a victim row), the DRAM chip asserts a back-off signal~\cite{saroiu2024ddr5, jedec2024jesd795c, canpolat2025chronus,canpolat2024understanding}, which forces the memory controller to issue a command called RFM~\cite{jedec2020ddr5}. The DRAM chip preventively refreshes potential victim rows upon receiving an RFM command.\footnote{{We refer the reader to recent works~\cite{saroiu2024ddr5, jedec2024jesd795c, canpolat2024breakhammer, canpolat2025chronus, canpolat2024understanding, hassan2024self} for more detail \om{on PRAC}.}}
}

\noindent\textbf{{Key Challenge: Updating Multiple Counters in PRAC.}}
{A SiMRA operation activates multiple rows \emph{simultaneously} and thus requires updating \omcr{0}{\emph{multiple}} PRAC row activation counters simultaneously to prevent \simra{}-induced read disturbance bitflips. However, since PRAC is designed to update \emph{only} one counter upon an \texttt{ACT} command (i.e., the one corresponding to the activated row), we need to modify the counter organization and operation in PRAC to support updating multiple counters. We assume a PRAC {implementation} described in Panopticon~\cite{bennett2021panopticon} where the counters reside in different subarrays than the rows that store data.}\footnote{{If counters are placed in the same subarray as the simultaneously activated DRAM rows, the counters that are simultaneously activated lose their values as \simra{} is a destructive operation that overwrites all activated rows with the result of the analog majority operation. Therefore, PRAC becomes insecure.}}
{We provide new methods to properly update multiple counters in PRAC: 1) an area-optimized solution and 2) a performance-optimized solution.}

\noindent\textbf{{\om{Area-Optimized Solution: PRAC-AO.}}}
{In the area-optimized solution \om{(PRAC-AO)}, we sequentially update the activation counters of each simultaneously activated row (i.e., update counters one by one). Doing so requires only one incrementer circuitry \omcr{0}{in the DRAM chip} to update counters and a small subarray (or mat) where row activation counters are kept (separately from the simultaneously activated rows), as proposed by Panopticon. However, since we need to update multiple (e.g., up to 32) counters, the latency for updating all counters becomes significantly higher than the standard memory access latency ($\trc{}$, typically \SI{46}{\nano\second}\om{-\SI{50}{\nano\second}}). For example, if the \simra{} operation activates 32 rows simultaneously, the mitigation mechanism needs to update 32 counters, which translates to a latency of 32$\times\trc{}$ (approximately \om{\SI{1.5}{\micro\second}-\SI{1.6}{\micro\second}}). 
}

{
\om{PRAC-AO likely induces prohibitive performance overheads due to the significant counter update latency. If we employ this solution as the read disturbance mitigation mechanism, 1) \simra{} operation throughput likely reduces significantly, and 2) every \simra{} operation blocks the target DRAM bank for $\approx$\SI{1.5}{\micro\second} during which the memory controller cannot serve any memory request targeting the same bank. \agyi{Such performance degradation would defeat the purpose of using PuD operations. To avoid this}, we propose \agyi{and evaluate} a 
performance-optimized solution (PRAC-PO) that does \emph{not} suffer from the drawbacks of the area-optimized solution.}
}

\noindent\textbf{{\om{Performance-Optimized~Solution:~PRAC-PO.}}}
{
We \omcr{0}{\emph{simultaneously}} update the activation counters of all simultaneously activated rows (i.e., update all counters at once). By doing so, we can alleviate the significant counter update latency of the area-optimized solution (e.g., \om{reducing} 32$\times\trc{}$ \om{to $\trc{}$}).
Assuming we can activate up to N rows simultaneously, this solution requires 1)~\agyi{simultaneous access to N different counter values,}
and 2)~N incrementer circuits to update all the counters simultaneously. 
We leave \agyi{a} detailed area overhead evaluation of this technique \omcr{0}{to} future work.

\noindent\textbf{{\om{Weighted Counting} Optimization.}} {\om{Due to prohibitive performance overheads of PRAC-AO as discussed in the PRAC-PO section, we optimize and focus our analysis on PRAC-PO.}
\om{PRAC-PO} securely prevents all read disturbance bitflips when configured for an RDT of $\approx$20 to account for SiMRA-induced read disturbance failures. However, this is a conservative configuration of our performance-optimized solution: the configuration induces prohibitive system performance overheads (as shown in~\figref{fig:rh_mit}). This is because the solution does \emph{not} take the heterogeneity in the read disturbance effects of RowHammer, CoMRA operations, and SiMRA operations. We find that the lowest \gls{hcfirst} \agyi{values} for RowHammer, \comra{}, and \simra{} \agy{are} $\approx$4K, $\approx$400, and $\approx$20, respectively.}
{Leveraging this heterogeneity,
we propose an optimization to reduce the performance overheads of PRAC: weighted contribution of different row activation types. \om{We assign each operation (i.e., RowHammer, \simra{}, and \comra{}) a weight equal to the lowest \gls{hcfirst} for RowHammer divided by that for the operation (e.g., 4K/20 = 200 for \simra{} and 4K/400 = 10 for \comra{}). Thus, we count}
 each \simra{} operation as 200 hammers (e.g., by adding 200 to the counters of simultaneously activated rows) and each \comra{} operation as 10 hammers to each row that participates in the \simra{} or \comra{} operation where 1 hammer is an activation of a DRAM row.}

\noindent\textbf{{Evaluation Methodology.}} 
{To evaluate performance, we conduct cycle-level simulations using Ramulator 2.0~\cite{ramulator2github,luo2023ramulator2} \omcr{0}{(new version of Ramulator~\cite{ramulatorgithub,kim2016ramulator})} with a realistic baseline system configuration.\footnote{{4.2GHz 5-core system, dual-rank DDR5 DRAM, FR-FCFS+Cap of 4~\cite{mutlu2007stall}. We simulate each workload until \omcr{0}{every} core execute 100M instructions.}} 
We extend Ramulator 2.0 to support \simra{} and \comra{} operations and update multiple \om{(up to 32)} counters simultaneously for PRAC\om{-PO}. 
We execute 60 five-core multiprogrammed workload mixes. Each mix is composed of four workloads from \ieycr{0}{five} major benchmark suites~\cite{spec2006, spec2017, tpc, fritts2009media,ycsb} and one synthetic workload that periodically performs \om{back-to-back} one \simra{} \om{with 32-row activation} and one \comra{} operation every N ns (where N \omcr{0}{ranges} from 125ns to 16$\mu$$s$).\footnote{{We vary the period to show how performance changes with the workload's \gls{pud} operation intensity.}}
We evaluate system performance using the weighted speedup metric~\cite{eyerman2008systemlevel,snavely00}.}

{We \om{analyze} two \om{different} PRAC\om{-PO} \om{implementations} and \om{evaluate} their overheads on the system performance: 1) \om{\emph{PRAC-PO-Naive}:} a naive \om{PRAC-PO implementation} without \om{weighted counting} optimization where the RowHammer threshold is reduced to lowest observed \gls{hcfirst} for \simra{} \om{(i.e., 20)} and 2) \om{\emph{PRAC-PO-Weighted Counting (PRAC-PO-WC)}: PRAC-PO with weighted counting} optimization.}

\noindent\textbf{{\om{Results.}}}
{
\figref{fig:rh_mit} presents the performance overheads of the \om{evaluated two PRAC\om{-PO} implementations, i.e., PRAC-PO-Naive and PRAC-PO-WC,} across 60 five-core multiprogrammed workload mixes. The x-axis shows the synthetic \gls{pud} workload's period of performing one \simra{} and one \comra{} operation (lower period indicates higher \gls{pud} operation intensity), and the y-axis shows \om{normalized} system performance \om{(higher is better)} in terms of weighted speedup normalized to a baseline with \emph{no} read disturbance mitigation.
}

\begin{figure}[ht]
    \centering
    \includegraphics[width=\linewidth]{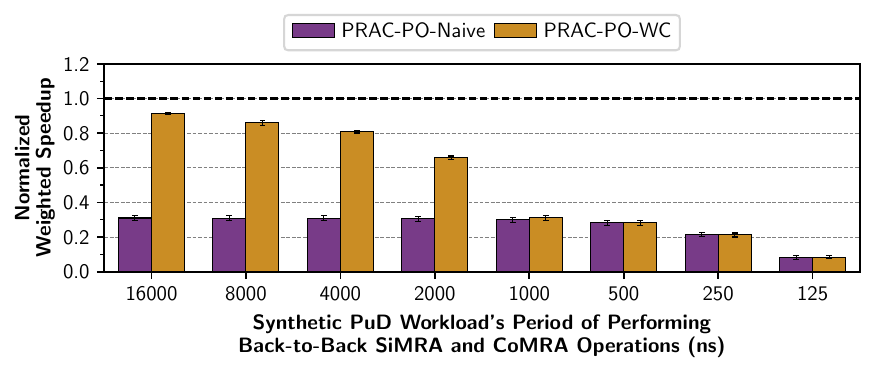}
    \caption{{\om{Performance impact of evaluated PRAC-PO implementations on 60 five-core multiprogrammed workloads.}}}
    \label{fig:rh_mit}
\end{figure}

{
 We make two major observations from \figref{fig:rh_mit}. 
 First, \om{even with our proposed optimizations, PRAC-PO-WC} incurs non-negligible performance overhead to mitigate PuDHammer-based read disturbance bitflips. We observe that PRAC-PO-WC incurs an average (maximum) system performance overhead of 48.26\% (98.83\%) across all tested \ieycr{0}{\gls{pud} operation intensity rates} and workloads.
 Second, across all evaluated \ieycr{0}{\gls{pud} operation intensity rates}, PRAC-PO-WC outperforms PRAC-PO-Naive. For example, at \omcr{0}{a} period of 4$\mu$$s$, PRAC-PO-WC induces 19.26\% average system performance overhead, whereas PRAC-PO-Naive induces 69.15\%.}

{We conclude \omcr{0}{that} \om{even with our proposed optimizations}, PRAC-PO-WC incurs significant performance overheads at high \gls{pud} operation intensity and non-negligible performance overheads at low \gls{pud} operation intensity to mitigate PuDHammer. We expect future work to introduce new efficient and effective mitigation mechanisms, as it has been happening analogously for RowHammer and RowPress.
}

\section{Related Work}
\label{sec:related_work}
We present the first experimental characterization of the read disturbance vulnerability caused by \omcr{0}{both} consecutive multiple-row activation (\comra{}) and simultaneous multiple-row activation (\simra{}) in real DDR4 DRAM chips. 

\noindent\textbf{Read Disturbance Characterization.} Many works~\cite{orosa2021deeper,kim2020revisiting,kim2014flipping, yaglikci2022understanding, lim2017active, park2016statistical, park2016experiments, ryu2017overcoming, yun2018study, lim2018study, luo2023rowpress, lang2023blaster, yaglikci2024svard, nam2024dramscope, olgun2023hbm, olgun2025variable, luo2024experimental, he2023whistleblower, luo2025revisiting, tugrul2025understanding} experimentally demonstrate (using real DDR3, DDR4, LPDDR4, and HBM2 DRAM chips) how a DRAM chip's read disturbance vulnerability varies with 1)~DRAM refresh rate~\cite{hassan2021utrr,frigo2020trrespass,kim2014flipping}, 2)~the physical distance between aggressor and victim rows~\cite{kim2014flipping,kim2020revisiting,lang2023blaster}, 3)~DRAM generation and technology node~\cite{orosa2021deeper,kim2014flipping,kim2020revisiting,hassan2021utrr}, 4)~temperature~\cite{orosa2021deeper,park2016experiments}, 5)~the time the aggressor row stays active~\cite{orosa2021deeper,park2016experiments,olgun2023hbm,olgun2024read,luo2023rowpress,nam2024dramscope,nam2023xray}, ~6)~physical location of the victim DRAM cell~\cite{orosa2021deeper,olgun2023hbm,olgun2024read,yaglikci2024svard}, 7)~wordline voltage~\cite{yaglikci2022understanding}, and 8)~supply voltage~\cite{he2023whistleblower}. {Our results are significantly different from what is already well-established in prior read disturbance characterization works. We study the read disturbance effects of \emph{processing-using-DRAM} operations' \agyi{row} activation patterns, \agyi{which} are fundamentally different from those studied by prior read disturbance characterization works: all prior read disturbance characterization works~\cite{orosa2021deeper,kim2020revisiting,kim2014flipping, yaglikci2022understanding, lim2017active, park2016statistical, park2016experiments, ryu2017overcoming, yun2018study, lim2018study, luo2023rowpress, lang2023blaster, yaglikci2024svard, nam2024dramscope, olgun2023hbm, olgun2025variable, luo2024experimental, he2023whistleblower, luo2025revisiting, tugrul2025understanding} activate \emph{at most one} aggressor row \emph{at a given time}. In \agy{contrast,} \agy{this paper investigates}
the read disturbance effect of activating \textbf{multiple} aggressor rows \emph{simultaneously} (i.e., SiMRA) or in \emph{quick succession} (i.e., CoMRA).
Our results demonstrate that the read disturbance effects of these new patterns (i.e., SiMRA and CoMRA operations) are fundamentally more damaging.
\ieycr{0}{In this work, we show that t}hey can induce bitflips with \emph{only} 26 operations \ieycr{0}{(\figref{fig:simra_rh_ds})}, which take 1.48$\mu$$s$, whereas the best known RowHammer pattern induces bitflips with 4123 aggressor row activations \ieycr{0}{(\figref{fig:simra_rh_ds})}, which take\om{s} 210.27$\mu$$s$ \ieycr{0}{and the best tested RowPress pattern (in terms of minimum \gls{hcfirst}) induces bitflips with 37 aggressor row activations at \gls{taggon}=70.2$\mu$$s$ \ieycr{0}{(\figref{fig:simra_rh_ds})}, which takes 2597.4$\mu$$s$}.
\emph{No} prior read disturbance characterization study evaluate{s} {the read disturbance effect of} SiMRA and CoMRA access patterns.}

\noindent\textbf{Multiple-Row Activation-based \gls{pud} Operations in \gls{cots} Chips.}
Several prior works demonstrate bulk bitwise~\cite{seshadri2017ambit,seshadri2015fast,hajinazar2021simdram,deoliveira2024mimdram,oliveira2025proteus} and \omcr{0}{bulk}data copy operations~\cite{seshadri2013rowclone} in COTS DRAM \omcr{0}{chips} using multiple-row activation~\cite{olgun2021quac,gao2019computedram,gao2022frac,yaglikci2022hira,olgun2023dram,yuksel2024functionally,olgun2022pidram, mutlu2024memory}. 
ComputeDRAM~\cite{gao2019computedram} activates three rows simultaneously to perform three-input majority and two-input AND and OR operations, and \two{} demonstrates copying one row's content to another row in DDR3 chips. FracDRAM~\cite{gao2022frac} shows that a DRAM cell in DDR3 chips can store fractional values.
QUAC-TRNG~\cite{olgun2021quac} simultaneously activates four rows to generate true random numbers in DDR4 chips \ieycr{0}{and a recent work~\cite{mutlu2024memory,yuksel2025experimental} experimentally studies the simultaneous activation of 2, 8, 16, and 32 rows in a subarray in COTS DDR4 DRAM chips}. HiRA~\cite{yaglikci2022hira} demonstrates that DDR4 chips can activate two rows in quick succession in electrically isolated subarrays.
DRAM Bender~\cite{olgun2023dram} demonstrates two{-}input AND and OR operations in DDR4 chips. PiDRAM~\cite{olgun2022pidram} provides an FPGA-based framework that enables real system studies of PuD techniques (e.g., RowClone). We do not use PiDRAM because PiDRAM is not designed to test DRAM chips but instead to evaluate PuD applications running end-to-end in a computing system. PiDRAM also performs SiMRA and CoMRA operations inside the DRAM chip, and thus it also suffers from the read disturbance effect that we investigate.
Prior work~\cite{yuksel2024functionally} demonstrates NOT and up to 16-input AND, NAND, OR, and NOR operations by simultaneously activating up to 48 rows in neighboring subarrays. 
A recent work~\cite{yuksel2024simultaneous} demonstrates up to 9-input majority operations and copying one row's content to up to 31 other rows concurrently by simultaneously activating up to 32 rows. \omcr{0}{N}one of these works investigates the read disturbance effects of such \omcr{0}{multiple-row activation} operations.

\section{Conclusion}
\label{sec:conclusion}
We presented our extensive characterization study on the interaction between read disturbance and multiple-row activation-based \omcr{0}{Processing-using-DRAM} operations in \nCHIPS{} \gls{cots} DDR4 DRAM chips from four major manufacturers. Our study leads to \param{26} new empirical observations and shares \param{9} key takeaway lessons, which demonstrate that multiple-row activation significantly exacerbates the DRAM read disturbance vulnerability \omcr{0}{either by itself} or \omcr{0}{when} combined with RowHammer, and this vulnerability gets worse \omcr{0}{under} various operating conditions and parameters (e.g., data pattern). We discuss \param{four} countermeasures against exploiting the read disturbance effect of \gls{pud} for future \gls{pud}-enabled systems. {We hope and expect that our detailed characterization results motivate and guide both 1)~system-level and architectural solutions to enable read-disturbance-resilient PuD systems and 2)~silicon-level works in understanding the underlying physical phenomena that explain PuDHammer’s characteristics.}

\section*{Acknowledgments}
We thank the anonymous reviewers of ISCA 2025 for their feedback.  
We thank the SAFARI Research Group members for providing a stimulating intellectual and scientific environment. We acknowledge the generous gifts from our industrial partners, including Google, Huawei, Intel, and Microsoft. This work, along with our broader work in Processing-in-Memory and memory systems, \ieycr{0}{is supported in part by the Semiconductor Research Corporation (SRC), the ETH Future Computing Laboratory (EFCL), AI Chip Center for Emerging Smart Systems (ACCESS), sponsored by InnoHK funding, Hong Kong SAR, European Union’s Horizon programme for research and innovation [101047160 - BioPIM], a Google Security and Privacy Research Award, and the Microsoft Swiss Joint Research Center.}

\balance
\bibliographystyle{IEEEtran}
\bibliography{refs,modules_ref}

\onecolumn
\appendix
\section{Tested DRAM Modules}
\label{sec:appendix_tested_dram_modules}
\newcommand*{\myalign}[2]{\multicolumn{1}{#1}{#2}}
\newcommand{\dimmid}[2]{\begin{tabular}[l]{@{}l@{}}#1~\cite{#1}\\#2~\cite{#2}\end{tabular}}

Table~\ref{tab:detailed_info} shows the characteristics of the DDR4 DRAM modules we test and analyze. We provide {the} module and chip identifiers, manufacturing date (Mfr. Date), {chip} density (Chip Den.), die revision {(Die Rev.)}, chip organization {(Chip Org.), and the subarray size} of tested DRAM modules. We report the manufacturing date of these modules in the form of $week-year$. Table~\ref{tab:detailed_info} shows the minimum and average \gls{hcfirst} values for double-sided RowHammer, CoMRA, and SiMRA across all tested rows.

\begin{table*}[ht]

\footnotesize
\centering
\caption{Characteristics of the tested DDR4 DRAM modules.}
\label{tab:detailed_info}
\resizebox{1.01\textwidth}{!}
{
\begin{tabular}{|l|l|l|c||cccc|ccc|}
\hline
\textbf{Module} & \textbf{Chip} & \textbf{Module Identifier} & \textbf{\#Modules} & \textbf{Mfr. Date} & \textbf{Chip} & \textbf{Die} & \textbf{Chip} & \multicolumn{3}{c|}{\textbf{Minimum (Average) }\pmb{\hcfirst{}}}  \\ 
\textbf{Vendor}&\textbf{Vendor}&\textbf{Chip Identifier}& \textbf{(\#Chips)} &\textbf{ww-yy}&\textbf{Den.}&\textbf{Rev.}&\textbf{Org.}&\textbf{RowHammer}&\textbf{CoMRA}&\textbf{SiMRA}\\
\hline
\hline
TimeTec & SK Hynix & \dimmid{75TT21NUS1R8-4}{H5AN4G8NAFR-TFC} & 1 (8) &  N/A & 4Gb & A & $\times8$ & 38.45K (112K) & 447 (5.84K) & 585 (6.62K) \\
\cline{1-11}
SK Hynix & SK Hynix & {\dimmid{HMA81GU7AFR8N-UH}{H5AN8G8NAFR-UHC}} & 8 (64) &  43-18 & 8Gb & A & $\times8$ & 25.0K (63.24K) & 1885 (45.28K) & 26 (16.14K) \\
\cline{1-11}
Kingston & SK Hynix & {\dimmid{KSM26ES8/16HC}{H5ANAG8NCJR-XNC}} & 2 (16) &  52-23 & 16Gb & C & $\times8$ & 6.25K (17.13K) & 4.54K (12.27K) & 48 (16.02K) \\
\cline{1-11}
SK Hynix & SK Hynix & {\dimmid{HMA81GU7DJR8N-WM}{H5AN8G8NDJR-WMC}} & 6 (48) &  N/A & 8Gb & D & $\times8$ & 7.58K (23.11K) & 632 (16.42K) & 95 (22.81K) \\
\cline{1-11}
\hline
\hline
Kingston & Micron & \dimmid{KVR21S15S8/4}{MT40A512M8RH-083E:B} & 1 (8) &  12-17 & 4Gb & B & $\times8$ & 126K (338K) & 93K (295K) & N/A \\
\cline{1-11}
Micron & Micron & {\dimmid{MTA4ATF1G64HZ-3G2E1}{MT40A1G16KD-062E:E}} & 4 (32) &  46-20 & 16Gb & E & $\times16$ & 4.89K (10.01K) & 3.72K (7.69K) & N/A \\
\cline{1-11}
Micron & Micron & {\dimmid{MTA18ASF4G72HZ-3G2F1}{MT40A2G8SA-062E:F}} & 4 (32) &  37-22 & 16Gb & F & $\times8$ & 4123 (9.03K) & 3.49K (7.06K) & N/A \\
\cline{1-11}
Micron & Micron & {\dimmid{KSM32ES8/8MR}{MT40A1G8SA-062E:R}} & 2 (16) &  12-24 & 8Gb & R & $\times8$ & 3.84K (9.32K) & 3.67K (7.67K) & N/A \\
\cline{1-11}
\hline
\hline
Samsung & Samsung & \dimmid{M378A2G43AB3-CWE}{K4AAG085WA-BCWE} & 1 (8) &  12-22 & 16Gb & A & $\times8$ & 6.70K (14.80K) & 5.26K (10.61K) & N/A \\ 
\cline{1-11}
Samsung & Samsung & {\begin{tabular}[l]{@{}l@{}}M391A2G43BB2-CWE~\cite{M391A2G43BB2-CWE}\\Unknown\end{tabular}} & 5 (40) &  15-23 & 16Gb & B & $\times8$ & 6.15K (14.79K) & 1875 (10.64K) & N/A \\ 
\cline{1-11}
Samsung & Samsung & \begin{tabular}[l]{@{}l@{}}M471A5244CB0-CRC~\cite{M471A5244CB0-CRC}\\Unknown\end{tabular} & 1 (4) &  19-19 & 4Gb & C & $\times16$ & 8.94K (25.83K) & 6.25K (18.40K) & N/A \\ 
\cline{1-11}
Samsung & Samsung & {\begin{tabular}[l]{@{}l@{}}M471A4G43CB1-CWE~\cite{M471A4G43CB1-CWE}\\Unknown\end{tabular}} & 1 (8) &  08-24 & 16Gb & C & $\times8$ & 6.81K (15.22K) & 4433 (10.95K) & N/A \\
\cline{1-11} 
Samsung & Samsung & {\dimmid{MTA4ATF1G64HZ-3G2B2}{MT40A1G16RC-062E:B}} & 1 (8) &  08-17 & 4Gb & E & $\times8$ & 15.77K (81.03K) & 11.72K (60.83K) & N/A \\ 
\cline{1-11}
\hline
\hline
Kingston & Nanya & {\begin{tabular}[l]{@{}l@{}}KVR24N17S8/8~\cite{KVR24N17S8/8}\\Unknown\end{tabular}}
 & 3 (24) &  46-20 & 8Gb & C & $\times8$ & 31.29K (128K) & 20.19K (107K) & N/A \\
\cline{1-11}
\hline 
\end{tabular}
}
\end{table*}

\section{{Discussion}}
\label{sec:ddr5}

\noindent\textbf{{PuDHammer on LPDDRx/DDR5.}}
{
We believe the fundamental observations of PuDHammer likely apply to LPDDRx/DDR5 as well (since the DRAM array essentially has the same organization \omcr{0}{as in DDR4}). Unfortunately, conducting LPDDRx/DDR5 experiments is extremely difficult since there is no robust open-source LPDDRx/DDR5 testing infrastructure.}
{As a result, we do not know 1) if we can perform SiMRA and/or CoMRA in COTS LPDDRx/DDR5 chips and 2) how \emph{severe} the read disturbance effects of SiMRA and CoMRA are in COTS LPDDRx/DDR5 chips.}

\noindent\textbf{{Effect of DDR5's Smaller Refresh Window.}}
{
Assuming 1) we can perform SiMRA and CoMRA operations in real DDR5 chips and 2) SiMRA and CoMRA's read disturbance effects in DDR5 and DDR4 are equally severe, DDR5's smaller refresh window (\SI{32}{\milli\second}) would \emph{not} prevent SiMRA or CoMRA bitflips. This is because the lowest \hcfirst{} observed for SiMRA (CoMRA) is 26 (447). Performing 26 (447) SiMRA (CoMRA) operations take 1.48$\mu$$s$ (42.24$\mu$$s$), a \om{very} small fraction of the \SI{32}{\milli\second} DDR5 refresh window.}

\end{document}